\pdfoutput=1

\documentclass[12pt,a4paper]{article}

\usepackage{ifthen} 
\usepackage{multirow} 
\newboolean{pdflatex}
\setboolean{pdflatex}{true} 

\newboolean{articletitles}
\setboolean{articletitles}{true} 

\newboolean{uprightparticles}
\setboolean{uprightparticles}{false} 

\newboolean{inbibliography}
\setboolean{inbibliography}{false} 

\usepackage[top=1in, bottom=1.25in, left=1in, right=1in]{geometry}

\usepackage{microtype}
\usepackage{lineno}  
\usepackage{xspace} 
\usepackage{caption} 

\usepackage{graphicx}  
\usepackage{color}
\usepackage{colortbl}
\graphicspath{{./figs/}} 
\usepackage{subfigure}

\usepackage{amsmath} 
\usepackage{amssymb}
\usepackage{amsfonts}
\usepackage{upgreek} 
\usepackage{pifont}
\newcommand*\patchAmsMathEnvironmentForLineno[1]{%
\expandafter\let\csname old#1\expandafter\endcsname\csname #1\endcsname
\expandafter\let\csname oldend#1\expandafter\endcsname\csname
end#1\endcsname
 \renewenvironment{#1}%
   {\linenomath\csname old#1\endcsname}%
   {\csname oldend#1\endcsname\endlinenomath}%
}
\newcommand*\patchBothAmsMathEnvironmentsForLineno[1]{%
  \patchAmsMathEnvironmentForLineno{#1}%
  \patchAmsMathEnvironmentForLineno{#1*}%
}
\AtBeginDocument{%
\patchBothAmsMathEnvironmentsForLineno{equation}%
\patchBothAmsMathEnvironmentsForLineno{align}%
\patchBothAmsMathEnvironmentsForLineno{flalign}%
\patchBothAmsMathEnvironmentsForLineno{alignat}%
\patchBothAmsMathEnvironmentsForLineno{gather}%
\patchBothAmsMathEnvironmentsForLineno{multline}%
\patchBothAmsMathEnvironmentsForLineno{eqnarray}%
}

\usepackage{hyperref}    
\usepackage[all]{hypcap} 

\def\lhcb {\mbox{LHCb}\xspace}

\def\MagUp {\mbox{\em Mag\kern -0.05em Up}\xspace}

\ifthenelse{\boolean{uprightparticles}}%
{

 \def\Ppi         {\ensuremath{\uppi}\xspace}

 \def\Ppsi        {\ensuremath{\uppsi}\xspace}

 \def\PDelta      {\ensuremath{\Delta}\xspace}                 
 \def\PXi      {\ensuremath{\Xi}\xspace}                 
 \def\PLambda      {\ensuremath{\Lambda}\xspace}                 
 \def\PSigma      {\ensuremath{\Sigma}\xspace}                 
 \def\POmega      {\ensuremath{\Omega}\xspace}                 
 \def\PUpsilon      {\ensuremath{\Upsilon}\xspace}                 
 
 \def\PB      {\ensuremath{\mathrm{B}}\xspace}                 
                  
 \def\PD      {\ensuremath{\mathrm{D}}\xspace}

 \def\PJ      {\ensuremath{\mathrm{J}}\xspace}                 
 \def\PK      {\ensuremath{\mathrm{K}}\xspace}

 \def\Pb      {\ensuremath{\mathrm{b}}\xspace}                 
 \def\Pc      {\ensuremath{\mathrm{c}}\xspace}                 
 \def\Pd      {\ensuremath{\mathrm{d}}\xspace}

 \def\Pi      {\ensuremath{\mathrm{i}}\xspace}

 \def\Pp      {\ensuremath{\mathrm{p}}\xspace}

 \def\Ps      {\ensuremath{\mathrm{s}}\xspace}                 
                  
 \def\Pu      {\ensuremath{\mathrm{u}}\xspace}

}
{

 \def\Ppi         {\ensuremath{\pi}\xspace}

 \def\Ppsi        {\ensuremath{\psi}\xspace}                 
                  
 \mathchardef\PDelta="7101
 \mathchardef\PXi="7104
 \mathchardef\PLambda="7103
 \mathchardef\PSigma="7106
 \mathchardef\POmega="710A
 \mathchardef\PUpsilon="7107
                  
 \def\PB      {\ensuremath{B}\xspace}                 
                  
 \def\PD      {\ensuremath{D}\xspace}

 \def\PJ      {\ensuremath{J}\xspace}                 
 \def\PK      {\ensuremath{K}\xspace}

 \def\Pb      {\ensuremath{b}\xspace}                 
 \def\Pc      {\ensuremath{c}\xspace}                 
 \def\Pd      {\ensuremath{d}\xspace}

 \def\Pi      {\ensuremath{i}\xspace}

 \def\Pp      {\ensuremath{p}\xspace}

 \def\Ps      {\ensuremath{s}\xspace}                 
                  
 \def\Pu      {\ensuremath{u}\xspace}

}

\makeatletter
\ifcase \@ptsize \relax
  \newcommand{\miniscule}{\@setfontsize\miniscule{4}{5}}
\or
  \newcommand{\miniscule}{\@setfontsize\miniscule{5}{6}}
\or
  \newcommand{\miniscule}{\@setfontsize\miniscule{5}{6}}
\fi
\makeatother

\DeclareRobustCommand{\optbar}[1]{\shortstack{{\miniscule (\rule[.5ex]{1.25em}{.18mm})}
  \\ [-.7ex] $#1$}}

\def\uquark    {{\ensuremath{\Pu}}\xspace}
\def\uquarkbar {{\ensuremath{\overline \uquark}}\xspace}

\def\dquark    {{\ensuremath{\Pd}}\xspace}
\def\dquarkbar {{\ensuremath{\overline \dquark}}\xspace}

\def\squark    {{\ensuremath{\Ps}}\xspace}

\def\cquark    {{\ensuremath{\Pc}}\xspace}

\def\bquark    {{\ensuremath{\Pb}}\xspace}
\def\bquarkbar {{\ensuremath{\overline \bquark}}\xspace}

\def\pion   {{\ensuremath{\Ppi}}\xspace}

\def\pip    {{\ensuremath{\pion^+}}\xspace}
\def\pim    {{\ensuremath{\pion^-}}\xspace}

\def\kaon    {{\ensuremath{\PK}}\xspace}
  \def\Kbar    {{\kern 0.2em\overline{\kern -0.2em \PK}{}}\xspace}

\def\KorKbar    {\kern 0.18em\optbar{\kern -0.18em K}{}\xspace}

\def\Kzb     {{\ensuremath{\Kbar{}^0}}\xspace}
\def\Kp      {{\ensuremath{\kaon^+}}\xspace}
\def\Km      {{\ensuremath{\kaon^-}}\xspace}

\def\KS      {{\ensuremath{\kaon^0_{\rm\scriptscriptstyle S}}}\xspace}

  \def\Dbar    {{\kern 0.2em\overline{\kern -0.2em \PD}{}}\xspace}
\def\D       {{\ensuremath{\PD}}\xspace}

\def\DorDbar    {\kern 0.18em\optbar{\kern -0.18em D}{}\xspace}

\def\Dm      {{\ensuremath{\D^-}}\xspace}

\def\B       {{\ensuremath{\PB}}\xspace}
\def\Bbar    {{\ensuremath{\kern 0.18em\overline{\kern -0.18em \PB}{}}}\xspace}

\def\BorBbar    {\kern 0.18em\optbar{\kern -0.18em B}{}\xspace}

\def\Bu      {{\ensuremath{\B^+}}\xspace}

\def\Bp      {{\ensuremath{\Bu}}\xspace}

\def\Bd      {{\ensuremath{\B^0}}\xspace}
\def\Bs      {{\ensuremath{\B^0_\squark}}\xspace}

\def\Bdb     {{\ensuremath{\Bbar{}^0}}\xspace}

\def\jpsi     {{\ensuremath{{\PJ\mskip -3mu/\mskip -2mu\Ppsi\mskip 2mu}}}\xspace}

  \def\Y#1S{\ensuremath{\PUpsilon{(#1S)}}\xspace}

\def\proton      {{\ensuremath{\Pp}}\xspace}
\def\antiproton  {{\ensuremath{\overline \proton}}\xspace}

\def\Lbar        {{\ensuremath{\kern 0.1em\overline{\kern -0.1em\PLambda}}}\xspace}
\def\LorLbar    {\kern 0.18em\optbar{\kern -0.18em \PLambda}{}\xspace}

\newcommand{\decay}[2]{\ensuremath{#1\!\to #2}\xspace}         

\def\to                 {\ensuremath{\rightarrow}\xspace}

\def\CP                {{\ensuremath{C\!P}}\xspace}

\newcommand{\phis}{{\ensuremath{\phi_{\squark}}}\xspace}

\def\SScomb     {{\ensuremath{\rm SScomb}}\xspace}
\def\SSpi       {{\ensuremath{\rm SS}\pion}\xspace}
\def\SSp        {{\ensuremath{\rm SS}\proton}\xspace}

\def\BdDpi        {\decay{\Bd}{D^-\pi^+}}

\def \BdDK        {\decay{\Bd}{D^-K^+}}

\def\BdToJPsiKS   {\decay{\Bd}{\jpsi\KS}}

\def\BdToKpi      {\decay{\Bd}{\Kp\pim}}

\def\AT#1     {\ensuremath{A_{\mathrm{T}}^{#1}}\xspace}           

\def\C#1      {\ensuremath{\mathcal{C}_{#1}}\xspace}                       
\def\Cp#1     {\ensuremath{\mathcal{C}_{#1}^{'}}\xspace}                    
\def\Ceff#1   {\ensuremath{\mathcal{C}_{#1}^{\mathrm{(eff)}}}\xspace}        
\def\Cpeff#1  {\ensuremath{\mathcal{C}_{#1}^{'\mathrm{(eff)}}}\xspace}       
\def\Ope#1    {\ensuremath{\mathcal{O}_{#1}}\xspace}                       
\def\Opep#1   {\ensuremath{\mathcal{O}_{#1}^{'}}\xspace}                    

\newcommand{\tev}{\ifthenelse{\boolean{inbibliography}}{\ensuremath{~T\kern -0.05em eV}\xspace}{\ensuremath{\mathrm{\,Te\kern -0.1em V}}}\xspace}
\newcommand{\gev}{\ensuremath{\mathrm{\,Ge\kern -0.1em V}}\xspace}
\newcommand{\mev}{\ensuremath{\mathrm{\,Me\kern -0.1em V}}\xspace}
\newcommand{\kev}{\ensuremath{\mathrm{\,ke\kern -0.1em V}}\xspace}
\newcommand{\ev}{\ensuremath{\mathrm{\,e\kern -0.1em V}}\xspace}
\newcommand{\gevc}{\ensuremath{{\mathrm{\,Ge\kern -0.1em V\!/}c}}\xspace}
\newcommand{\mevc}{\ensuremath{{\mathrm{\,Me\kern -0.1em V\!/}c}}\xspace}
\newcommand{\gevcc}{\ensuremath{{\mathrm{\,Ge\kern -0.1em V\!/}c^2}}\xspace}
\newcommand{\gevgevcccc}{\ensuremath{{\mathrm{\,Ge\kern -0.1em V^2\!/}c^4}}\xspace}
\newcommand{\mevcc}{\ensuremath{{\mathrm{\,Me\kern -0.1em V\!/}c^2}}\xspace}

\def\mum  {\ensuremath{{\,\upmu\rm m}}\xspace}

\def\invfb   {\ensuremath{\mbox{\,fb}^{-1}}\xspace}

\def\ps   {\ensuremath{{\rm \,ps}}\xspace}
\def\fs   {\ensuremath{\rm \,fs}\xspace}

\def\gsim{{~\raise.15em\hbox{$>$}\kern-.85em
          \lower.35em\hbox{$\sim$}~}\xspace}
\def\lsim{{~\raise.15em\hbox{$<$}\kern-.85em
          \lower.35em\hbox{$\sim$}~}\xspace}

\def\sPlot{\mbox{\em sPlot}}

\def\sqs   {\ensuremath{\protect\sqrt{s}}\xspace}
\def\pp       {\mbox{$pp$}\xspace}

\def\ptot       {\mbox{$p$}\xspace}
\def\pt         {\mbox{$p_{\rm T}$}\xspace}

\def\mrad{\ensuremath{\rm \,mrad}\xspace}

\def\evtgen     {\mbox{\textsc{EvtGen}}\xspace}

\def\geant      {\mbox{\textsc{Geant4}}\xspace}

\def\photos     {\mbox{\textsc{Photos}}\xspace}

\def\pythia     {\mbox{\textsc{Pythia}}\xspace}

\def\tell1  {TELL1\xspace}
\def\ukl1   {UKL1\xspace}

\usepackage{cite} 
\usepackage{mciteplus}

\usepackage{longtable} 

\begin{document}

\renewcommand{\thefootnote}{\fnsymbol{footnote}}
\setcounter{footnote}{1}


\begin{titlepage}
\pagenumbering{roman}

\vspace*{-1.5cm}
\centerline{\large EUROPEAN ORGANIZATION FOR NUCLEAR RESEARCH (CERN)}
\vspace*{1.5cm}
\noindent
\begin{tabular*}{\linewidth}{lc@{\extracolsep{\fill}}r@{\extracolsep{0pt}}}
\ifthenelse{\boolean{pdflatex}}
{\vspace*{-1.5cm}\mbox{\!\!\!\includegraphics[width=.14\textwidth]{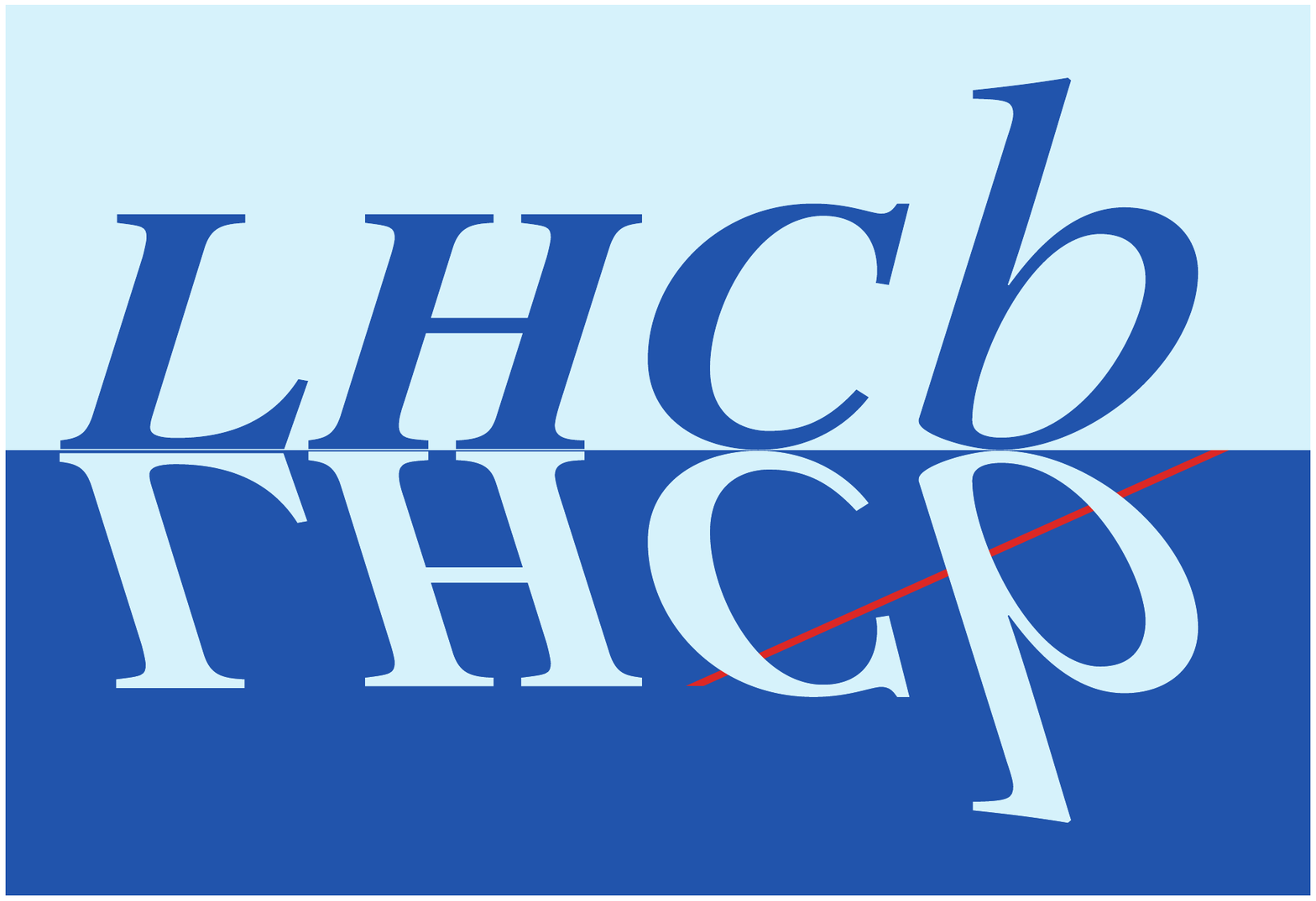}} & &}%
{\vspace*{-1.2cm}\mbox{\!\!\!\includegraphics[width=.14\textwidth]{lhcb-logo.eps}} & &}%
\\
 & & CERN-EP-2016-251 \\  
 & & LHCb-PAPER-2016-039 \\  
 & & 19 October 2016 \\ 
 & & \\
\end{tabular*}
\vspace*{2.0cm}

{\bf\boldmath\huge
\begin{center}
 New algorithms for identifying the flavour of \Bd mesons using pions and protons
\end{center}
}
\vspace*{2.0cm}
\begin{center}
The LHCb collaboration\footnote{Authors are listed at the end of this paper.}
\end{center}
\vspace{\fill}
\begin{abstract}
\noindent
Two new algorithms for use in the analysis of \pp collision are developed
to identify the flavour of \Bd mesons at production
using pions and protons from the hadronization process. 
The algorithms are optimized and calibrated on data, using
\BdDpi decays from \pp collision data collected by \lhcb 
at centre-of-mass energies of 7 and 8\tev.
The tagging power of the new pion algorithm is 60\% greater than the previously 
available one; the algorithm using protons to identify the flavour of a 
\Bd meson is the first of its kind.  
\end{abstract}
\vspace*{2.0cm}
\begin{center}
Published in EPJC
\end{center}
\vspace{\fill}
{\footnotesize 
\centerline{\copyright~CERN on behalf of the \lhcb collaboration, licence \href{http://creativecommons.org/licenses/by/4.0/}{CC-BY-4.0}.}}
\vspace*{2mm}

\end{titlepage}


\newpage
\setcounter{page}{2}
\mbox{~}

\cleardoublepage


\renewcommand{\thefootnote}{\arabic{footnote}}
\setcounter{footnote}{0}


\pagestyle{plain} 
\setcounter{page}{1}
\pagenumbering{arabic}


\section{Introduction}
\label{sec:Introduction}
Violation of \CP symmetry in the \B system was observed for the first time in the interference
between mixing and decay processes~\cite{MixPDG2014}. Any measurement of a decay-time-dependent asymmetry
requires the determination of the flavour of the \B meson at production.
For \B mesons produced in \pp collisions, this information is obtained by 
means of several flavour-tagging algorithms that exploit the correlations 
between \B flavour and other particles in the event.

Algorithms determining the flavour content of \B meson by using particles 
associated to its production are called same-side (SS) taggers.
As an example, in the production of \Bd mesons from 
excited charged \B mesons decaying via strong interaction to $\Bd \pi^+$,
the pion charge identifies the 
initial flavour of the \Bd meson.\footnote{The inclusion of charge-conjugate
processes is implied throughout the paper, unless otherwise noted.}
A charge correlation can also arise from the hadronization process of the 
\bquark quark. 
When a \bquarkbar and a \dquark quark hadronize as a \Bd meson,  
it is likely that the corresponding \dquarkbar quark ends up in a charged 
pion (\uquark\dquarkbar), or in an antiproton 
($\uquarkbar\hspace{0.1em}\uquarkbar\hspace{0.1em}\dquarkbar$). 
The \Bd meson and the pion or antiproton are produced in nearby regions of 
phase space.
Other algorithms used at LHCb, called opposite-side 
(OS) taggers~\cite{LHCb-PAPER-2011-027, LHCb-PAPER-2015-027},
attempt to identify the flavour of the other \bquark hadron
produced in the same event.

A simple cut-based SS algorithm selecting pions was 
successfully used by LHCb for tagging \BdToJPsiKS decays~\cite{LHCb-PAPER-2015-004} 
in the measurement of $\sin 2\beta$, and an SS kaon 
tagger~\cite{LHCb-PAPER-2015-056} based on a neural network was used to 
determine the flavour of \Bs mesons in measurements of the \CP-violating phase 
\phis~\cite{LHCb-PAPER-2014-059, LHCb-PAPER-2014-019, LHCb-PAPER-2014-051}.
This paper presents two new SS algorithms exploiting the charge correlation of pions
and protons with \Bd mesons, denoted \SSpi and \SSp.
This is the first time that protons are used for flavour tagging.
The two algorithms are combined into a single tagger, \SScomb.
Both algorithms are based on multivariate selections and are 
optmized, calibrated and validated using \BdDpi and \BdToKpi decays 
collected by LHCb in Run 1.

The performance of a flavour-tagging algorithm is measured by its tagging 
efficiency $\varepsilon_{\mathrm{tag}}$, mistag fraction $\omega$, dilution $D$,
and tagging power $\varepsilon_{\mathrm{eff}}$, defined as
\begin{equation}
\varepsilon_{\mathrm{tag}} = \frac{R+W}{R+W+U}, \ \ \ \ \
\omega = \frac{W}{R+W}, \ \ \
 D = 1-2\omega,\ \ \ \varepsilon_{\mathrm{eff}} = \varepsilon_{\mathrm{tag}} D^2,
\end{equation}
where $R$, $W$, and $U$ are the numbers of correctly-tagged, incorrectly-tagged,
and untagged \Bd signal candidates.
The tagging power determines the sensitivity to the measurement of a 
decay-time-dependent \CP asymmetry~\cite{BFactories}, as it quantifies the effective 
reduction in the sample size of flavour-tagged \Bd candidates. 
It is the figure of merit used to optimize the algorithms.
Each algorithm provides a decision on the flavour of the \Bd candidate and 
an estimate of the probability $\eta$ that this decision is incorrect. 
The probability is used to determine a weight applied to the \Bd candidate, in order 
to maximize the tagging power of a sample of \Bd mesons in a time-dependent analysis. 
The probabilities provided by the two SS taggers are used to combine their 
decisions into the \SScomb decision, 
which can be further combined with the decision of other
taggers~\cite{LHCb-PAPER-2011-027, LHCb-PAPER-2015-027}.

\begin{table}[tb]
 \centering
\caption[]{Expected correlation between the flavour of a \B meson and the
hadronization products.}
\begin{tabular}{c|ccc}
\B meson & pion & proton & kaon \\ 
\hline
\rule{0pt}{3ex}  \Bd & \pip & \antiproton & \Kzb \\
 \Bp  & \pim & \antiproton & \Km \\
\end{tabular}\label{tab:rightTag}
\end{table}

The expected relationship between the flavour of charged and neutral \B mesons 
and the charge of the tagging particle is reported in Table~\ref{tab:rightTag}.
For a \Bu meson the same correlation as for a \Bd meson holds in the case of 
protons, but with opposite charge in the case of pions. 
In addition, the tagging kaons carry the same charge as pions, while they
are neutral for a \Bd. 
Since misidentified hadrons affect the tagging efficiency and 
the mistag fraction of charged and neutral mesons in different ways,
\Bu decays cannot be reliably used for the tuning and calibration 
of the SS taggers.
As a consequence, \Bd decays are used, and a time-dependent analysis 
is required to determine the mistag fraction.

\section{Detector}
\label{sec:detector}
The LHCb detector~\cite{Alves:2008zz, LHCb-DP-2014-002} 
is a single-arm forward spectrometer covering the \mbox{pseudorapidity} 
range $2<\eta <5$,
designed for the study of particles containing \bquark or \cquark
quarks. The detector includes a high-precision tracking system
consisting of a silicon-strip vertex detector surrounding the $pp$
interaction region, a large-area silicon-strip detector located
upstream of a dipole magnet with a bending power of about
$4{\rm\,Tm}$, and three stations of silicon-strip detectors and straw
drift tubes placed downstream of the magnet.
Regular reversal of the magnet polarity allows a quantitative assessment of
detector-induced charge asymmetries.
The tracking system provides a measurement of momentum, \ptot, of charged 
particles with a relative uncertainty that varies from 0.5\% at low momentum 
to 1.0\% at 200\gevc.
The minimum distance of a track to a primary vertex (PV), the impact parameter
(IP), is measured with a resolution of $(15+29/\pt)\mum$,
where \pt is the component of the momentum transverse to the beam, in\,\gevc.

Particularly relevant for this analysis is the identification of the different 
species of charged hadrons, which mainly relies on the information
of two ring-imaging Cherenkov detectors.
The first one covers the low and intermediate momentum region 2 - 40\gevc over the 
full spectrometer angular acceptance of 25 - 300\mrad. 
The second Cherenkov detector covers the high momentum region 
15 - 100\gevc over the angular range 15 - 120\mrad~\cite{LHCb-DP-2012-003}.

Photons, electrons and hadrons are identified by a calorimeter system 
consisting of scintillating-pad and preshower detectors, an electromagnetic
calorimeter and a hadronic calorimeter. Muons are identified by a
system composed of alternating layers of iron and multiwire
proportional chambers.
The online event selection is performed by a
trigger~\cite{LHCb-DP-2012-004}, 
which consists of a hardware stage, based on information from the calorimeter 
and muon systems, followed by a software stage, which applies a full event
reconstruction.
At the hardware trigger stage, events are required to have a muon with 
high \pt or a hadron, photon or electron with high transverse energy in the 
calorimeters. 
The software trigger requires a two-, three- or four-track
secondary vertex detached from the PV. 
A multivariate algorithm~\cite{BBDT} is used for
the identification of secondary vertices consistent with the decay
of a \bquark hadron.

Samples of simulated events are used to model the signal mass and 
decay-time distributions. In the simulation, $pp$ collisions are generated using \pythia~\cite{Sjostrand:2006za,*Sjostrand:2007gs}  
with a specific \lhcb configuration~\cite{LHCb-PROC-2010-056}.  
Decays of hadronic particles are described by \evtgen~\cite{Lange:2001uf}, 
in which final-state radiation is generated using \photos~\cite{Golonka:2005pn}. 
The interaction of the generated particles with the detector, and its response, 
are implemented using the \geant toolkit~\cite{Allison:2006ve, *Agostinelli:2002hh} 
as described in Ref.~\cite{LHCb-PROC-2011-006}.

\section{Development of the same-side taggers}
\label{sec:Tuning} 
The \SSpi and \SSp algorithms are developed following similar strategies.
A sample of \Bd mesons decaying into the flavour-specific final state $D^-\pi^+$, 
with \Dm candidates reconstructed in the final state \Kp \pim \pim, is selected 
using requirements similar to those presented in 
Ref.~\cite{LHCb-PAPER-2014-037}. The sample is collected from
\pp collisions at $\sqs=8\tev$, corresponding to an integrated luminosity of 
2\invfb. 
Tagging pion or proton candidates, with their charge correlated with the \Bd flavour, 
are selected by means of a set of loose selection requirements and a 
multivariate classifier, as described below. 
The \BdDpi candidates are separated randomly into three disjoint subsamples of equal size.  
The first sample is used for training the multivariate classifiers, the second is used for determining 
the probability of an incorrect tagging decision, and the third is used to evaluate 
the calibration of the mistag probability.  

The correctness of a tagging decision is evaluated by comparing the charge
of the tagging particle with the \Bd decay flavour as determined by the 
reconstructed final state.
Those \Bd candidates that have oscillated before decaying enter the training 
process with an incorrectly assigned production flavour.
In the training phase the dilution is reduced by requiring the decay time of 
the reconstructed \Bd mesons to be smaller than 2.2\ps. 
This value was optimized with 
simulated events and reduces the fraction of oscillated candidates 
to about 11\%, keeping 66\% of the original sample.

\begin{figure}[tb]
\centering
\includegraphics[width=0.5\textwidth]{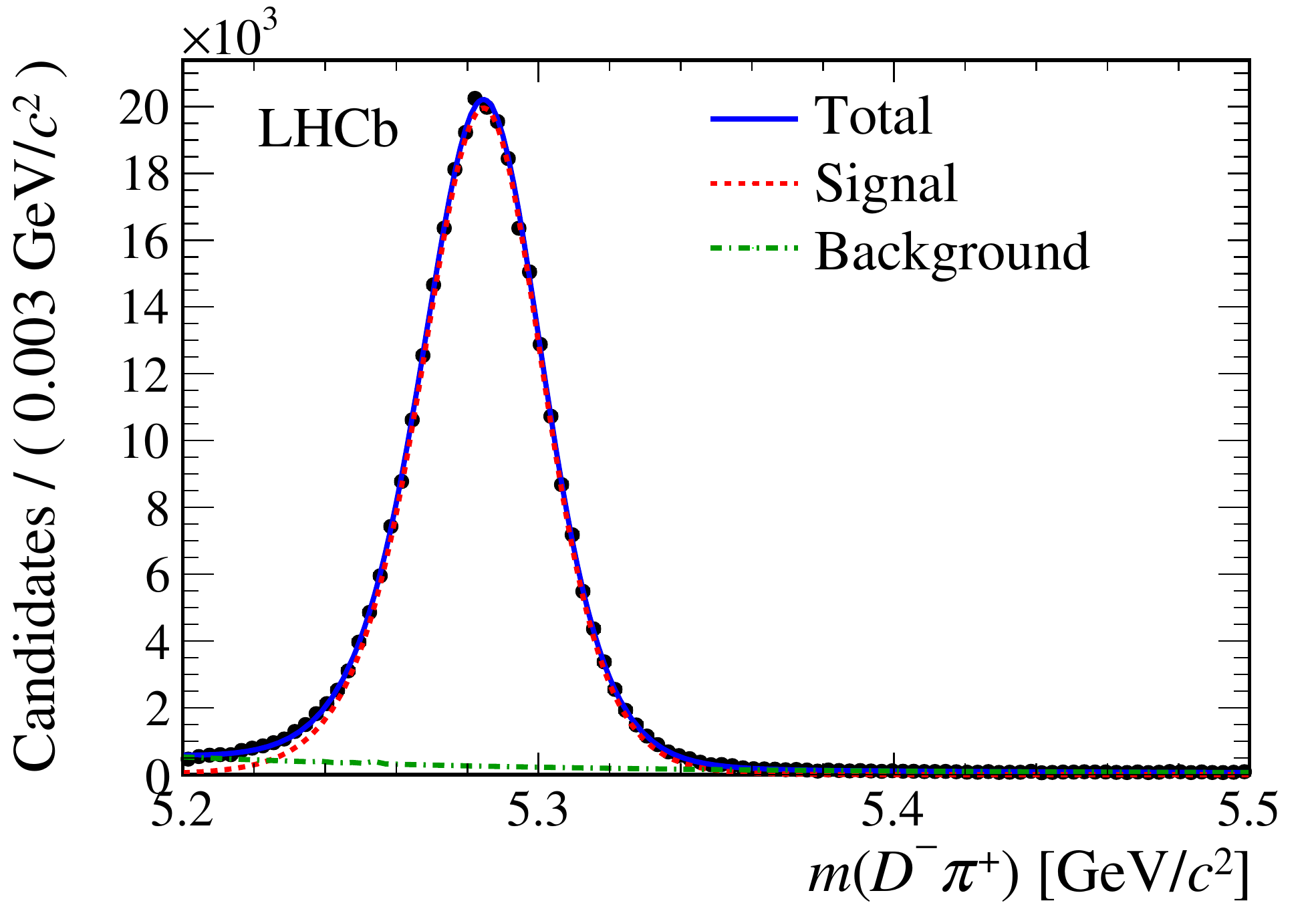}
\caption[]{Mass distribution of \BdDpi candidates with fit projections overlaid.
Data points (black dots) correspond to the \Bd candidates selected in the 
2 \invfb data sample collected at $\sqs=8 \tev$. 
The solid blue curve represents the total fit function which is the sum of signal 
(red dashed) and combinatorial background (green dash-dotted).}
\label{fig:Dpi_massfit}
\end{figure}

The signal and background components of the \Bd sample are determined by an 
unbinned maximum likelihood fit to 
the \Dm \pip mass distribution of the selected candidates in the region 
[5.2, 5.5]\gevcc.
The signal is described by a Johnson's $S_U$ distribution~\cite{JohnsonPDF}, while the combinatorial background is modelled by the sum of an exponential function 
and a constant. 
All parameters are free to vary in the fit.
A small component of \BdDK decays ($\sim$1.2\% as estimated from simulation), 
with the kaon misidentified as a pion, 
is neglected in the fit.
The number of signal candidates in the full 2\invfb sample, estimated by the mass 
fit and shown in Fig.~\ref{fig:Dpi_massfit}, is 300\,370 $\pm$ 674. 
The fit to the mass distribution is used to assign event-by-event weights 
(sWeights), using the \sPlot\ technique~\cite{Pivk:2004ty}. 
The weights are subsequently used to subtract 
the background contribution when training the \SSpi and \SSp classifiers 
and in the fits to the \Bd decay-time distribution. 

The loose selection requirements reduce the multiplicity of pion (proton) 
candidates to 2.3 (1.7) per \BdDpi signal candidate, 
and are reported in Table~\ref{tab:presel}.
Only tracks with hits in all tracking detectors are considered as tagging candidates.
The following observables are used:
the $\chi^2/\mathrm{ndf}$ of the track fit, where 
ndf is the number of degrees of freedom, 
the track transverse momentum $p_{\mathrm{T}}^{\mathrm{track}}$, 
the ratio between the track impact parameter with respect to the PV associated to the \Bd meson 
and the error on this variable $\mathrm{IP}/\sigma_{\mathrm{IP}}$, 
the ratio between the track impact parameter with respect to any other PV in 
the event and its error $\mathrm{IP}_{\mathrm{PU}}/\sigma_{\mathrm{IP}_{\mathrm{PU}}}$, 
the difference between the logarithms of the likelihood of the proton 
and pion hypothesis $\log L_p -\log L_{\pi}$, or 
kaon and pion hypothesis $\log L_K -\log L_{\pi}$.
The likelihoods for the various mass hypothesis are determined using the track 
and the Cherenkov angle information, as described in Ref.~\cite{LHCb-DP-2012-002}.
For particles passing the loose selection criteria                             
the efficiency to identify a pion is 89\% with a kaon misidentification probability 
of 2\%,
while the efficiency to identify a proton is 92\% with a pion
misidentification probability of 5\%.
Since mutually exclusive particle identification criteria are imposed,
a given tagging track is identified either as a pion or as a proton.
If more than one PV is reconstructed in the event, the PV associated to the 
\Bd meson is the one which has the smallest increase in the vertex-fit $\chi^2$ 
when adding the \Bd meson to the PV.

Additional requirements are introduced on the system formed by the tagging 
particle and the reconstructed \Bd meson. 
They are applied to the total transverse momentum of the system $p_{\mathrm{T}}^{\mathrm{tot}}$,
the difference between the pseudorapidity of the \Bd candidate and the tagging 
particle $\Delta \eta$, the azimuthal angle $\Delta \phi$ between the \Bd 
candidate and the tagging particle, and the difference between the invariant 
mass of the system and the mass of the \Bd and of the tagging particle 
$\Delta Q = m(\Bd+h)- m(\Bd) - m(h)$, where $h$ denotes the hadron, $\pi$ or $p$.
The vertex formed by the \Bd meson and the tagging particle is required to have
the $\chi^2$ of vertex fit $\chi^2_{B^0-\mathrm{track}}$, less than 100.

\begin{table}[tb]
\centering
\caption{Loose selection requirements for the \SSpi and \SSp algorithms. 
The variables used as input for the BDT classifiers are indicated by \checkmark.}
\begin{tabular}{c|cc|cc}
& \multicolumn{2}{c}{\SSpi} & \multicolumn{2}{c}{\SSp} \\
Variable & selection &  BDT & selection & BDT\\ \hline
$\chi^2_{\mathrm{track}}/\mathrm{ndf}$ & $< 3$ & \checkmark & $< 3$ & - \\ 
$p_{\mathrm{T}}^{\mathrm{track}}$ [GeV/c] & $> 0.4$ & \checkmark & $> 0.4$& \checkmark \\
$p^{\mathrm{track}}$ [GeV/c] & - & \checkmark & - & \checkmark \\
$\mathrm{IP}/\sigma_{\mathrm{IP}} $& $< 4$ & \checkmark & $< 4$ & \checkmark \\
$\mathrm{IP}_{\mathrm{PU}}/\sigma_{\mathrm{IP}_{\mathrm{PU}}}$ & $> 3$ &- & - & -\\
$\log L_p -\log L_{\pi}$ & $< 5$& - & $> 5$ & \checkmark \\
$\log L_K -\log L_{\pi}$ & $< 5$& \checkmark  & - & -\\
$p_{\mathrm{T}}^{\Bd}$ [GeV/c] &  -& \checkmark  & - & - \\
$p_{\mathrm{T}}^{\mathrm{tot}}$ [GeV/c] &  $> 3$& \checkmark  &$> 3$ & \checkmark \\
$\chi^2_{B^0-\mathrm{track}}$ & $< 100$ & - & $< 100$ & - \\
$\Delta Q$ [GeV/c$^2$]&$< 1.2$ & \checkmark &$< 1.3$ & \checkmark \\
$\Delta \eta$ & $< 1.2$ & \checkmark & $< 1.2$ & \checkmark \\
$\Delta \phi$ [rad] & $< 1.1$ & \checkmark &$< 1.2$ & -  \\
$\Delta R$ & - & \checkmark & - & \checkmark \\
$PV_{\mathrm{tracks}}$ & - & \checkmark & - & \checkmark \\
\end{tabular}
\label{tab:presel}
\end{table}

\begin{figure}[b]
\centering
{\includegraphics[width=0.49\textwidth]{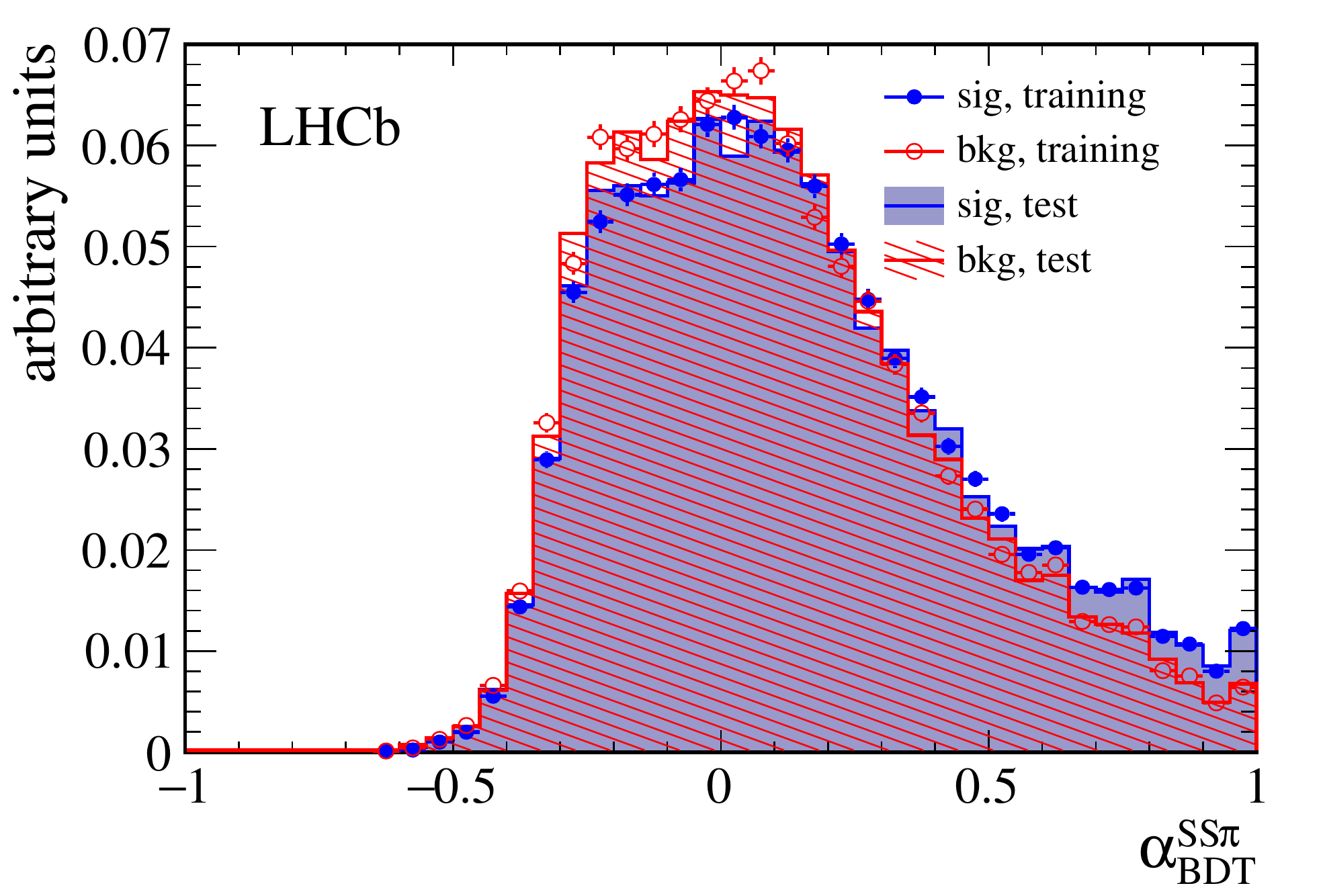}}
{\includegraphics[width=0.49\textwidth]{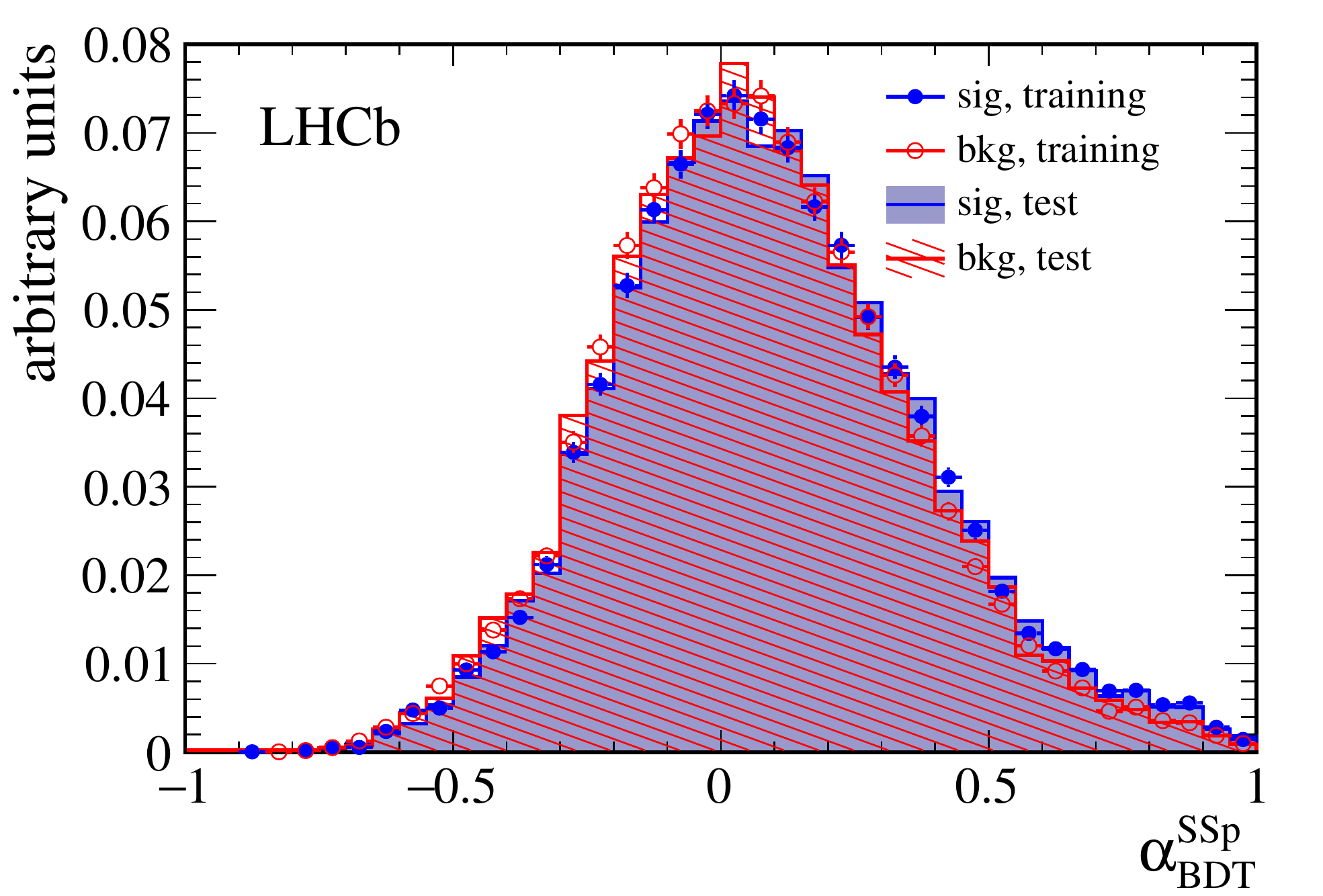}}
\caption{Distribution of the BDT output of signal (correct-tag decision) 
and background (wrong-tag decision) tagging particles, 
for (left) \SSpi and (right) \SSp taggers.
In case of multiple tagging candidates per \Bd candidate, only 
the candidate with the highest BDT output value is shown. 
}
\label{fig:pion_BDT_response_best}
\end{figure}

The multivariate classifiers used for the selection of the tagging particles 
are boosted decision trees (BDT)~\cite{Breiman}
using the AdaBoost~\cite{AdaBoost} method to enhance and to increase the
stability with respect to statistical fluctuations.
This choice has been shown to be optimal with respect to the achievable tagging power.
The  classifiers take most of the above observables as input, as specified in Table~\ref{tab:presel}.
In addition the BDTs use the following variables: 
the momentum of the tagging particle $p^{\mathrm{track}}$, 
the transverse momentum of the \Bd candidate $p_{\mathrm{T}}^{\Bd}$,
the separation of tagging particle and the \Bd candidate 
$\Delta R = \sqrt{\Delta \phi^2 + \Delta \eta^2}$, 
and the number of tracks contributing to the PV fit $PV_{\mathrm{tracks}}$. 
The sWeights are used to subtract the contribution of background \Bd candidates
in the training of the classifiers. 
The charge of the tagging particle determines the flavour of the \Bd candidate.
In case of multiple tagging particle candidates per \Bd candidate, 
the tagging particle with the highest BDT output value is chosen. 
The BDT outputs, $\alpha_{\mathrm{BDT}}$, are shown in 
Fig.~\ref{fig:pion_BDT_response_best}.
The global separation between signal and background is small, 
but enough to provide useful information to determine 
the flavour of the \Bd candidate, as shown below.

\section{Evaluation and calibration of mistag probability}
\label{sec:calibration} 

\subsection{The SS$\pi$ and SSp taggers}

The BDT output is transformed into an estimate of the mistag probability 
through linear regression.
The decay-time distribution of all tagged \Bd candidates is considered and
the dilution due to mixing is decoupled by means of a full time-dependent 
analysis.
Tagged \Bd candidates are divided into eight bins of the BDT output 
and for each bin the probability of an incorrect tagging
decision is determined from an unbinned maximum likelihood fit to the 
distribution of the measured decay time $t$ of the candidates, using the sWeights.
The probability density function (PDF) for the signal is described as
\begin{equation}
\label{eqn:SimpleTfit}
 {\cal S}(t,q) = {\cal N}~a(t)~e^{-t'/\tau_d} ( 1 + q(1-2\omega) 
\cos(\Delta m_d~t'))\otimes {\cal R}(t-t'),
\end{equation}
where $t'$ represents the true decay time, {$\cal N$} is a normalization factor,
$\omega$ is the average mistag fraction in the bin, $q$ is the mixing state 
($q=+1$ when the flavour at production and the flavour at decay are the same,
 $q=-1$ otherwise), ${\cal R}(t-t')$ is the decay-time resolution and $a(t)$ is the 
decay-time acceptance. The \Bd lifetime $\tau_d$, and the mixing frequency $\Delta m_d$,
are fixed in the fit to their known values~\cite{PDG2014}.

\begin{figure}[tb]
\centering
\includegraphics[width=0.49\textwidth]{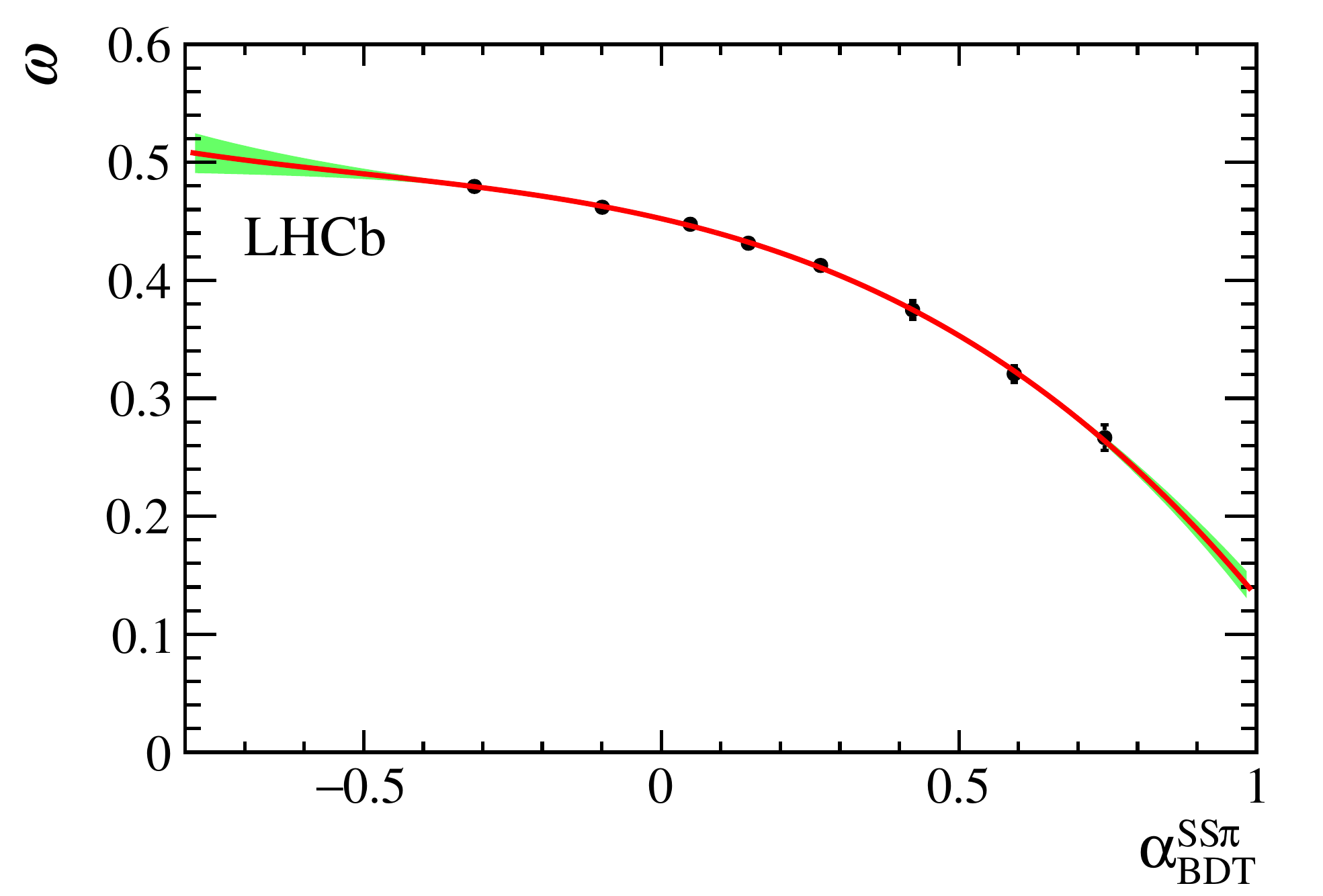}
\includegraphics[width=0.49\textwidth]{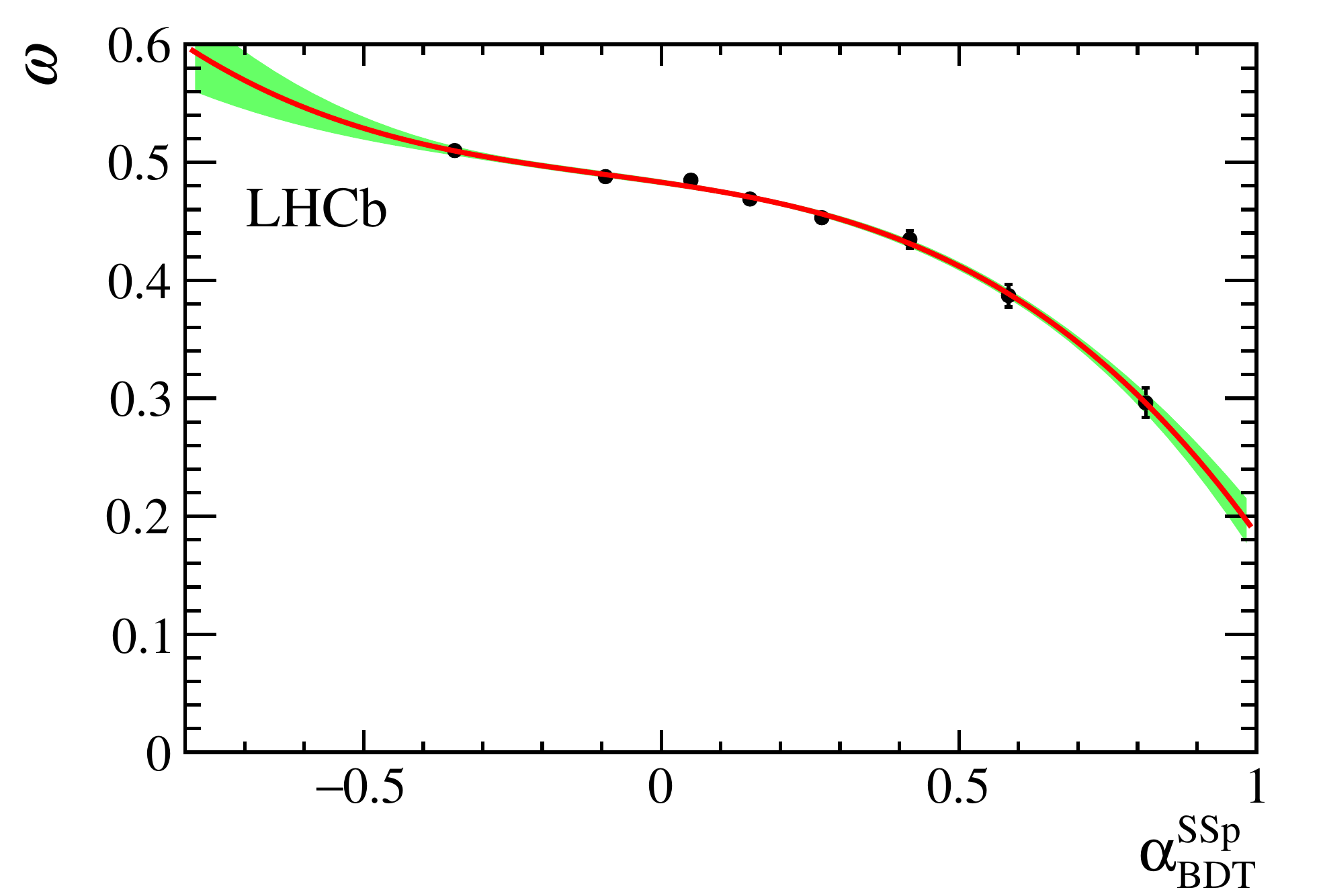}
\caption
{Measured average mistag fraction in bins of (left) \SSpi and (right) \SSp BDT output. 
The plots are obtained with the test sample of background-subtracted \BdDpi candidates.
The green-shaded area shows the confidence range within $\pm 1\sigma$.}
\label{fig:BDTcalibration}
\end{figure}

Equation~\ref{eqn:SimpleTfit} is obtained under the assumption of zero 
width difference $\Delta \Gamma_d$ and neglecting the production 
and detection asymmetries between \Bd and \Bdb. 
The decay-time resolution is modelled by a Gaussian function with a fixed 
width of 50\,\fs, as determined from simulation. 
The decay-time acceptance $a(t)$, is described by a parametric function based on 
cubic splines~\cite{splines} whose nodes have fixed position 
and whose parameters are determined from data.
Figure~\ref{fig:BDTcalibration} shows the measured average mistag rate per 
subsample, 
interpolated with a third-order polynomial that represents $\eta$ as a function 
of $\alpha_{\mathrm{BDT}}$, for the \SSpi and \SSp taggers.

This polynomial parametrization is then used to determine the mistag 
probability $\eta(\alpha_{\mathrm{BDT}})$ of a \Bd candidate. 
Tagging particles with $\eta(\alpha_{\mathrm{BDT}})>0.5$ are rejected.
With the third subsample of \Bd candidates, 
it is checked that the estimated mistag probability corresponds to the true
value by measuring the mistag fraction $\omega$ with an unbinned likelihood 
fit to the decay-time distribution of the \Bd candidates.
Possible differences between the mistag probability of \Bd and \Bdb mesons 
may arise from the different interaction cross-sections of hadrons and 
antihadrons in the detector material and from differences in detection 
efficiencies of positive and negative hadrons.
They are taken into account in the decay-time fit by defining the variables  
\begin{equation}
  \overline{\omega} = (\omega^{\Bd} + \omega^{\Bdb})/2, ~~~~~~~~~~
  \Delta \omega =  \omega^{\Bd} - \omega^{\Bdb}, \\
\label{eqn:OmegaDiff}
\end{equation}
where $\omega^{\Bd}$ and $\omega^{\Bdb}$ are the mistag fractions related 
to \Bd and \Bdb.
Assuming a linear relation between the measured and estimated mistag fractions,
the calibration functions are written as 
\begin{equation}
\begin{matrix}
\label{eqn:Omegacalib}
  \omega^{\Bd}(\eta) = p_0^{\Bd} + p_1^{\Bd}(\eta - \langle \eta \rangle), \\
  \omega^{\Bdb}(\eta) = p_0^{\Bdb} + p_1^{\Bdb}(\eta - \langle \eta \rangle),
\end{matrix}
\end{equation}
where $p_i^{\Bd}$ and $p_i^{\Bdb}$ (with  $i=0,1$) 
are the calibration parameters.
The average calibration parameters and the differences between the 
\Bd and \Bdb parameters are defined as 
\begin{equation}
\overline{p}_i = (p_i^{\Bd}  + p_i^{\Bdb})/2,  ~~~~~~~~~~~
\Delta p_i = p_i^{\Bd}  - p_i^{\Bdb}. 
\end{equation}
The use of the arithmetic mean $\langle \eta \rangle$ of the $\eta$ 
distribution aims at decorrelating $p_0$ and $p_1$. A perfect 
calibration corresponds to $\overline{p}_0=\langle \eta \rangle$ and $\overline{p}_1=1$.

A difference in the number of reconstructed and tagged \Bd and \Bdb mesons 
arises from several possible sources.
Two of these sources are considered in the fit by introducing an asymmetry in the detection 
efficiency of the final state particles, defined as
\begin{equation}
A_{\mathrm{det}} = \frac{\varepsilon_{\mathrm{det}}^{D^+\pi^-}-\varepsilon_{\mathrm{det}}^{D^-\pi^+}}
{\varepsilon_{\mathrm{det}}^{D^+\pi^-}+\varepsilon_{\mathrm{det}}^{D^-\pi^+}},
\end{equation}
and an asymmetry of the tagging efficiencies, defined as
\begin{equation}
A_{\mathrm{tag}} =  \frac{ \varepsilon_{\mathrm{tag}}^{\Bdb} - \varepsilon_{\mathrm{tag}}^{\Bd} }
{ \varepsilon_{\mathrm{tag}}^{\Bdb} +\varepsilon_{\mathrm{tag}}^{\Bd} }.
\end{equation}
With these additional inputs, the PDF becomes
\begin{equation}
\label{eqn:Timefit}
 {\cal S}(t,q) = {\cal N}~a(t)~e^{-t'/\tau_d} (C_{\mathrm{cosh}} + C_{\mathrm{cos}} \cos(\Delta m_d~t'))\otimes {\cal R}(t-t').
\end{equation}
The coefficients $C_{\mathrm{cosh}}$ and $C_{\mathrm{cos}}$ are 
\begin{align}
C_{\mathrm{cosh}}=&(1-r~A_{\mathrm{det}})\Bigl(1-\frac{a_{\mathrm{sl}}^d}{2}~\frac{1+r}{2}\Bigr)\Biggl((1+A_{\mathrm{prod}}+A_{\mathrm{tag}})\Bigl(\frac{1-d}{2}+d(\omega+\Delta\omega)\Bigr) \nonumber \\
& + (1-A_{\mathrm{prod}}-A_{\mathrm{tag}})\Bigl(\frac{1+d}{2}-d(\omega-\Delta \omega)\Bigr)\Bigl(1+\frac{a_{sl}^d}{2}\Bigr)\Biggr), \nonumber \\
C_{\mathrm{cos}}=&-r(1-r~A_{\mathrm{det}})\Bigl(1-\frac{a_{\mathrm{sl}}^d}{2}~\frac{1+r}{2}\Bigr)\Biggl((1+A_{\mathrm{prod}}+A_{\mathrm{tag}})\Bigl(\frac{1-d}{2}+d(\omega+\Delta\omega)\Bigr)\nonumber \\
& - (1-A_{\mathrm{prod}}-A_{\mathrm{tag}})\Bigl(\frac{1+d}{2}-d(\omega-\Delta \omega)\Bigr)\Bigl(1+\frac{a_{\mathrm{sl}}^d}{2}\Bigr)\Biggr),  \label{eq:Ccos}
\end{align}
where $r$ is the \B meson flavour at decay ($r=+1$ for 
$\Bd \rightarrow D^- \pi^+$, $r=-1$ for 
$\Bdb \rightarrow D^+ \pi^-$) and $d$ is the tagging decision ($d=+1$ for 
$\pi^+$ ($\overline{p}$), $d=-1$ for $\pi^-$ ($p$)). 
These coefficients also take into account the production asymmetry,
$A_{\mathrm{prod}} = \frac{N_{\Bdb}-N_{\Bd}}{N_{\Bdb}+N_{\Bd}}$,  
and the asymmetry in mixing, or flavour-specific asymmetry, 
$a_{\mathrm{sl}}^{d}$.
These two asymmetries cannot be distinguished from the tagging 
and detection asymmetries and are fixed in the fit.
The production asymmetry is fixed to the value measured
in Ref.~\cite{LHCb-PAPER-2014-042}, $A_{\mathrm{prod}}=(-0.58 \pm 0.70)\%$, 
while $a_{\mathrm{sl}}^{d}$ is fixed to the world average
$a_{\mathrm{sl}}^d = (-0.15 \pm 0.17)\%$~\cite{HFAG}.
The effect of their uncertainties on the calibration parameters
is included in the systematic uncertainty.

The calibration parameters for the two taggers obtained in the fit to 
the calibration sample of \BdDpi decays are reported in 
Table~\ref{tab:final_table}.
The correlations between the calibration parameters are below 10\%,
except for the asymmetry of the tagging efficiencies, which has a correlation of about 16\% 
with $\Delta p_0$ and $\Delta p_1$ and about 64\% with  $A_{\mathrm{det}}$.
For the \SSpi tagger,  $A_{\mathrm{tag}}$, $\Delta p_0$ and  $\Delta p_1$ are 
zero within one standard deviation, 
showing no significant difference in tagging behaviour between 
\Bd and \Bdb decays.
For the \SSp tagger, it is found that $\Delta p_0<0$, as a consequence of the higher 
interaction cross-section of anti-protons with matter compared to protons.
A similar effect is reported for kaon taggers~\cite{LHCb-PAPER-2015-056}.
The fit result of the detection asymmetry is comparable for the two
taggers ($A_{\mathrm{det}}^\SSpi=(-0.87 \pm 0.48)\%$, $A_{\mathrm{det}}^{{\mathrm{SS}}{\mathit p}}=
(-0.66 \pm 0.62)\%$) 
and in agreement with that found in Ref.~\cite{LHCb-PAPER-2014-053}.
The systematic uncertainties on the parameters will be described in 
Section~\ref{sec:syst}.

\begin{table}[tb]
\centering
\caption[]{Calibration parameters for the \SSpi, \SSp and \SScomb taggers 
where the first uncertainties are statistical and the second are systematic.}
\resizebox{\textwidth}{!}{
\begin{tabular}{cccc}
  &  \SSpi & \SSp & \SScomb \\
\hline
$\langle \eta \rangle$ & $\phantom{0}0.444$ & $\phantom{0}0.461$ & $\phantom{0}0.439$ \\
$\overline{p}_0$ & $\phantom{+}0.446\phantom{0} \pm 0.003\phantom{0} \pm 0.001\phantom{0}$ & $\phantom{+}0.468\phantom{0} \pm 0.004\phantom{0} \pm 0.001\phantom{0}$ & $\phantom{+}0.441\phantom{0} \pm 0.003\phantom{0} \pm 0.002\phantom{0}$ \\
$\overline{p}_1$ &  $\phantom{+}1.05\phantom{00} \pm 0.05\phantom{00} \pm 0.01\phantom{00}$ & $\phantom{+}1.04\phantom{00} \pm 0.08\phantom{00} \pm 0.02\phantom{00}$ & $\phantom{+}0.99\phantom{00} \pm 0.04\phantom{00} \pm 0.02\phantom{00}$ \\
$\Delta p_0$ & $-0.0028 \pm 0.0036 \pm 0.0016$ & $-0.0218 \pm 0.0048 \pm 0.0016$ & $-0.0056 \pm 0.0036 \pm 0.0018$ \\
$\Delta p_1$ & $\phantom{+}0.015\phantom{0} \pm 0.074\phantom{0} \pm 0.014\phantom{0}$ & $\phantom{+}0.140\phantom{0} \pm 0.112\phantom{0} \pm 0.019\phantom{0}$ & $\phantom{+}0.052\phantom{0} \pm 0.060\phantom{0} \pm 0.017\phantom{0}$ \\
$A_{\mathrm{tag}}$ & $-0.001\phantom{0} \pm 0.007\phantom{0} \pm 0.007\phantom{0}$ & $\phantom{+}0.008\phantom{0} \pm 0.009\phantom{0} \pm 0.007\phantom{0}$ & $\phantom{+}0.002\phantom{0} \pm 0.007\phantom{0} \pm 0.007\phantom{0}$ \\
\end{tabular}
}
\label{tab:final_table}
\end{table}                                                                    

After calibration, the total tagging power of the sample is calculated as
\begin{equation}
\label{eqn:tagging_power}
\varepsilon_{\mathrm{eff}} = \frac{\sum_{i=1}^{N_{\mathrm{tag}}}(1-2{\overline{\omega}}(\eta_i))^2s_i}
{\sum_{j=1}^{N} s_j}
\end{equation}
where $s_i$ is the sWeight of the candidate $i$, 
$N$ and $N_{\mathrm{tag}}$ are the numbers of total and tagged candidates, 
having mistag probability $\eta_i$, and  
the average mistag fraction ${\overline{\omega}}(\eta_i)$ is calculated using 
Eqs.~\ref{eqn:OmegaDiff} and~\ref{eqn:Omegacalib}.
Candidates with a mistag probability larger than 0.5 are considered untagged and are removed from 
the sum in the numerator, effectively setting $\omega(\eta_i) = 0.5$.
The tagging performances for the \SSpi and \SSp taggers 
are reported in Table~\ref{tab:TagPowerSS}.

  \begin{table}[tb]
 \centering
 \caption{Tagging efficiencies and tagging power of the \SSpi, \SSp and
  \SScomb algorithms. The \SScomb efficiencies are shown splitting
  the sample in candidates tagged exclusively by \SSpi or \SSp, or by both.
  As explained in the text, there is a large overlap between the \SSpi and \SSp taggers.}
 \begin{tabular}{cccc}
Tagger &  Sample &  $\varepsilon_{\mathrm{tag}}$ [$\%$] & $\varepsilon_{\mathrm{eff}}$ [$\%$] \\\hline
\SSpi && $71.96 \pm 0.23$ & $1.69 \pm 0.10$ \\\hline
\SSp  && $38.56 \pm 0.15$ & $0.53 \pm 0.05$ \\\hline
\multirow{4}{*}{\SScomb}&  \SSpi only   & $35.91 \pm 0.14$ & $0.95 \pm 0.08$ \\
  &\SSp  only    & $\phantom{0}8.75 \pm 0.10$ & $0.12 \pm 0.02$ \\
  &\SSpi\& \SSp  & $34.74 \pm 0.15$ & $1.04 \pm 0.07$ \\
  \cline{2-4}
  & total        & $79.40 \pm 0.23$ & $2.11 \pm 0.11$ \\
 \end{tabular}
 \label{tab:TagPowerSS}
  \end{table}

\begin{figure}[tb]
{\includegraphics[width=0.5\textwidth]{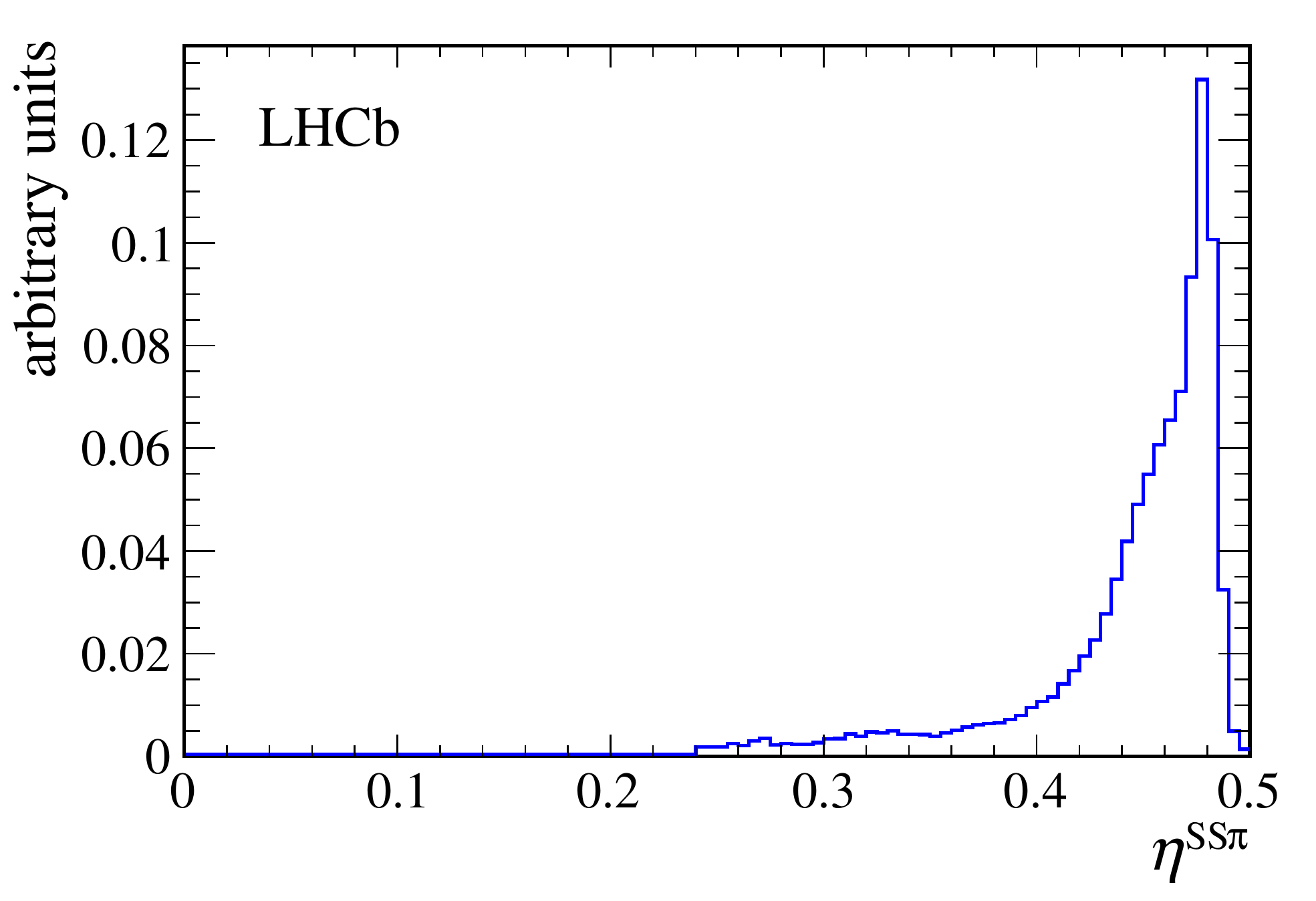}}
{\includegraphics[width=0.5\textwidth]{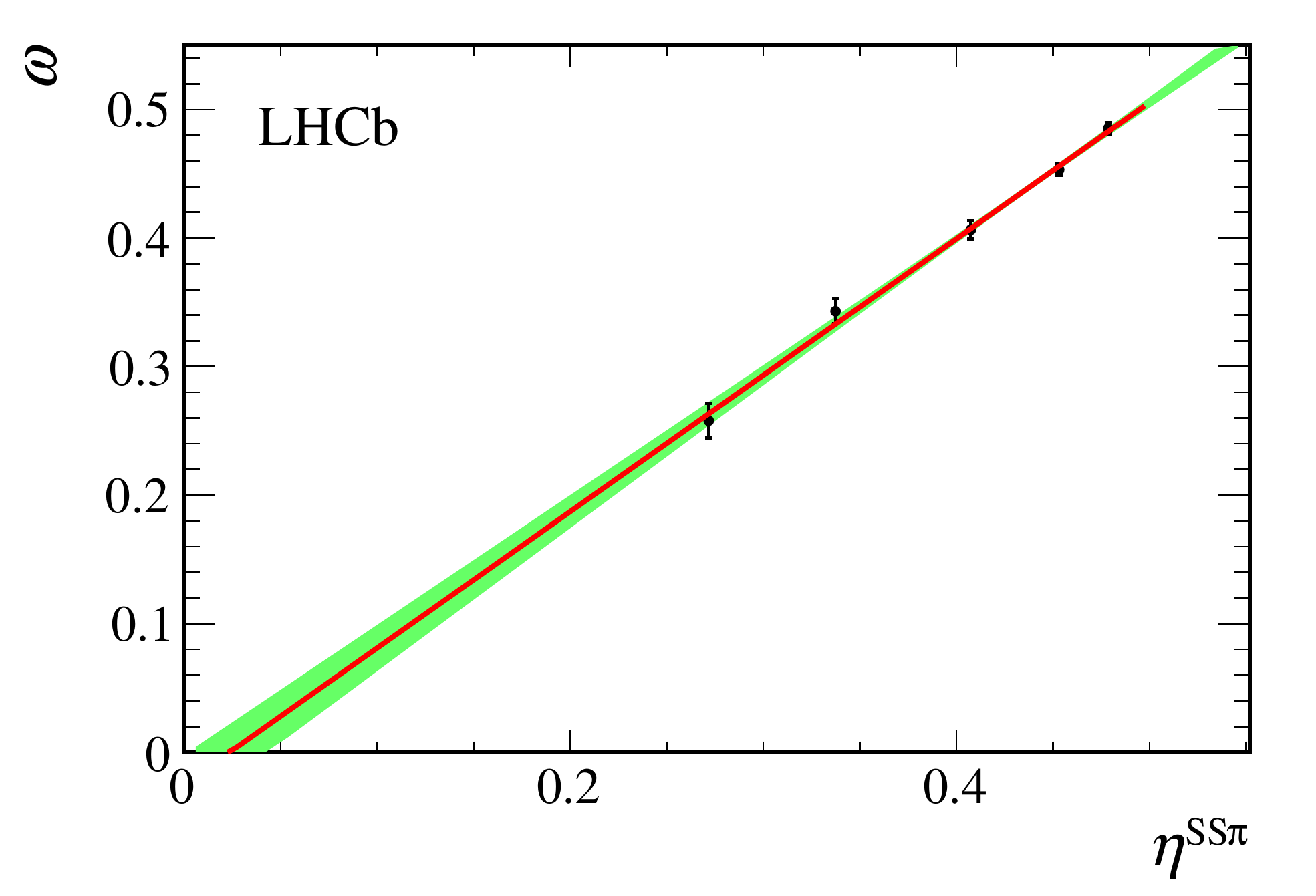}}\\
{\includegraphics[width=0.5\textwidth]{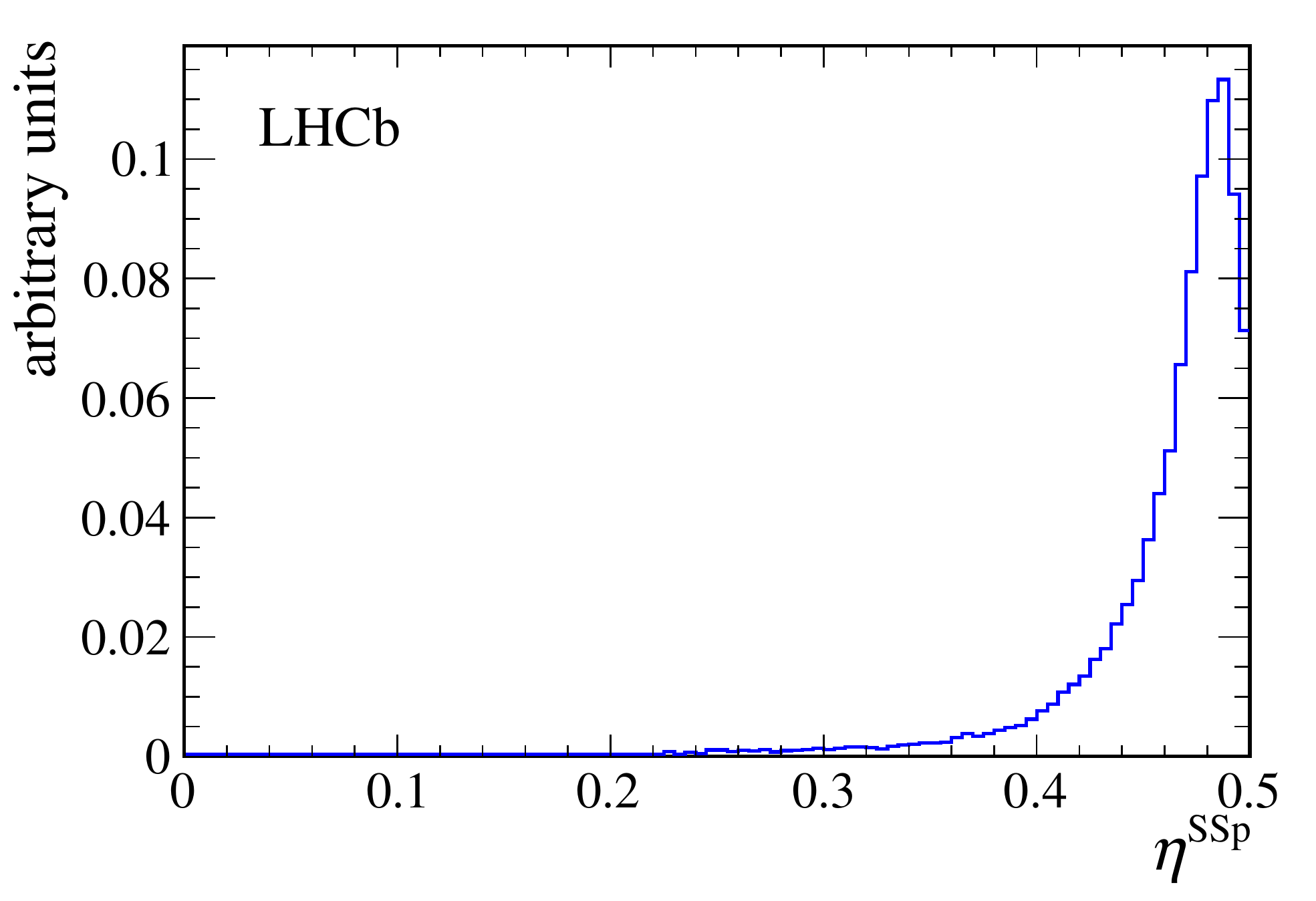}}
{\includegraphics[width=0.5\textwidth]{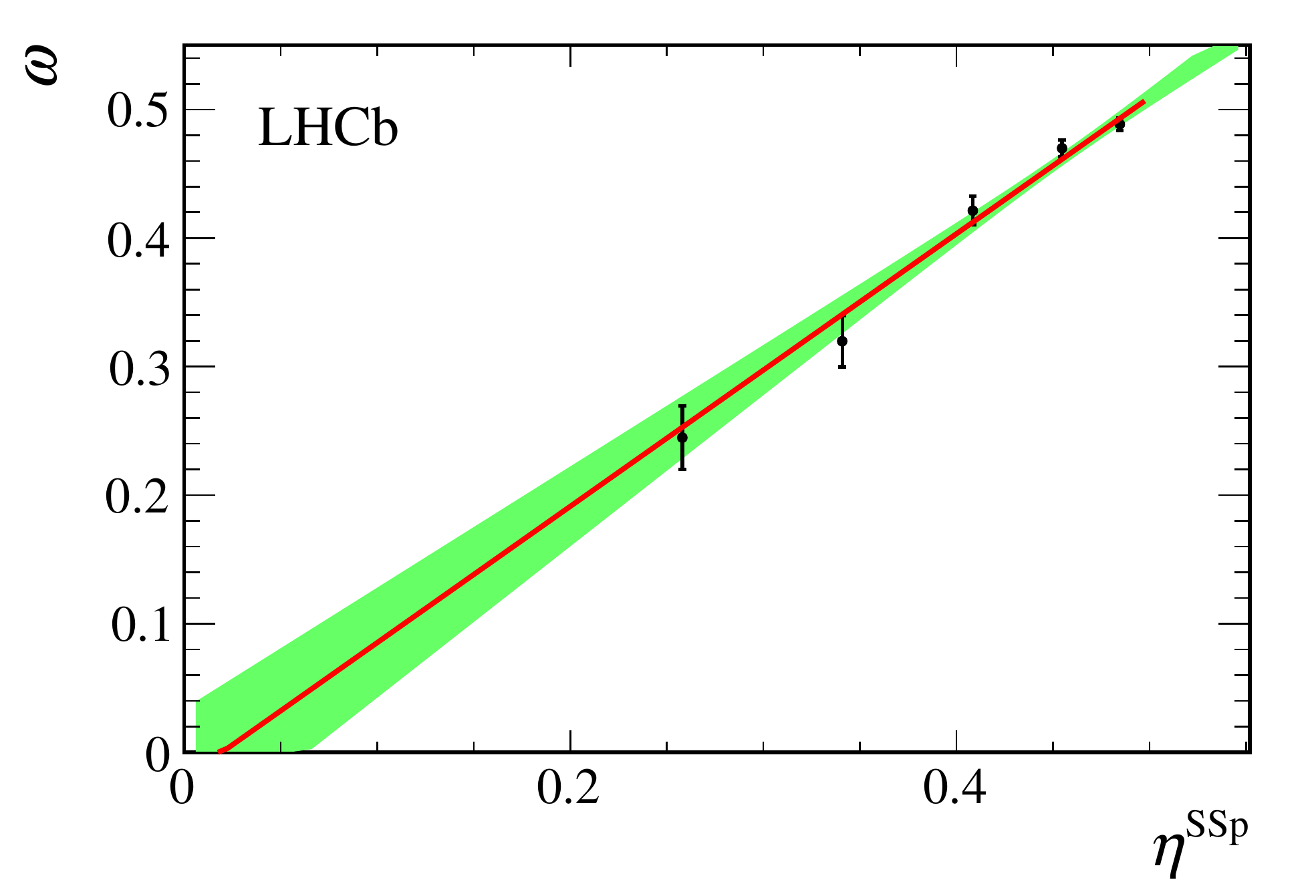}}
\caption{(top left) Distribution of the mistag probability 
$\eta^\SSpi$ and (top right) measured mistag fraction $\omega$ as a function
of $\eta^\SSpi$. 
(bottom left) Distribution of the mistag probability
$\eta^\SSp$ and (bottom right) measured mistag fraction $\omega$ 
as a function of $\eta^\SSp$. 
The green-shaded area shows the 68\% confidence range.}
\label{fig:etaSSpi}
\end{figure}

\begin{figure}[tb]
{\includegraphics[width=0.5\textwidth]{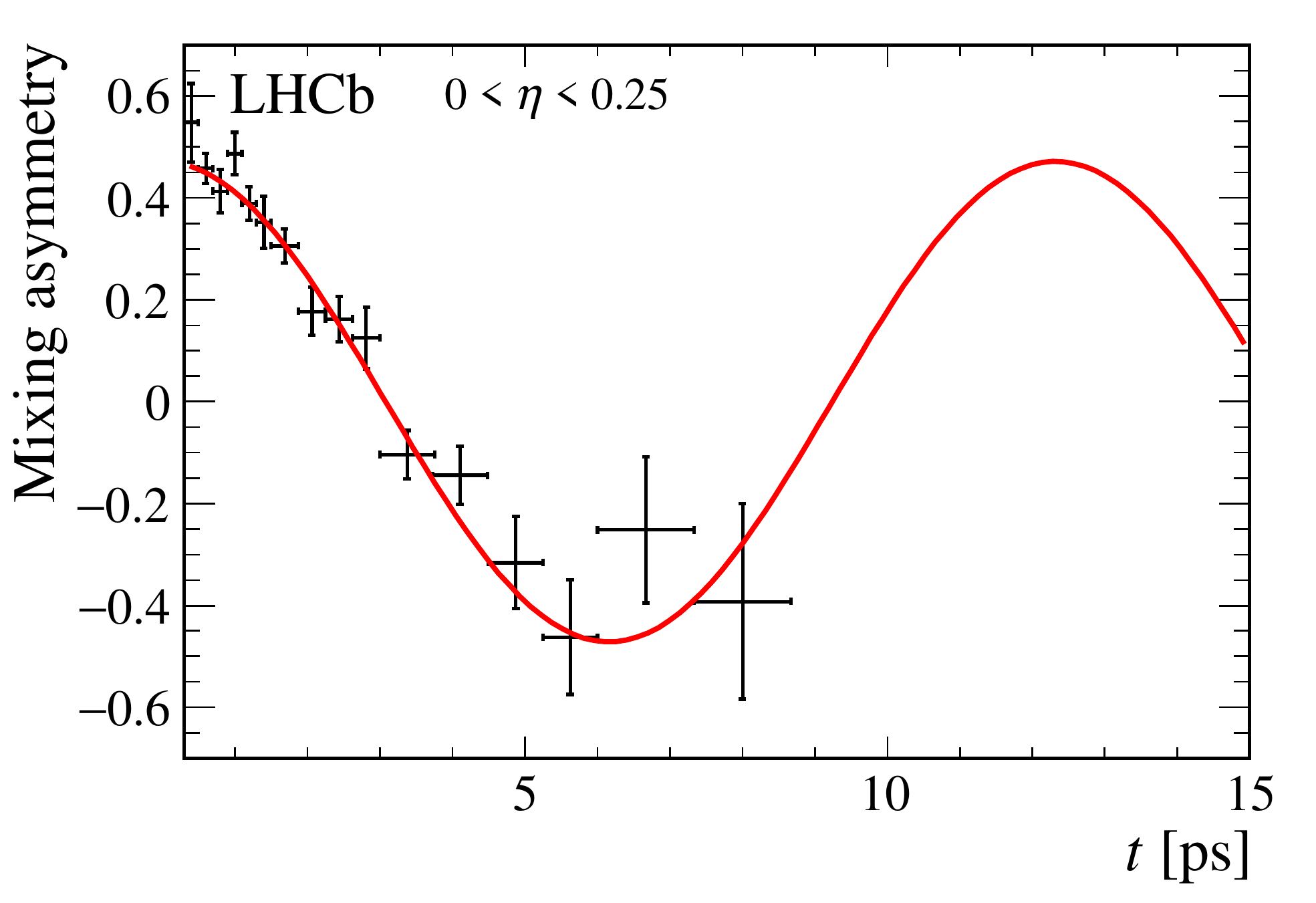}}
{\includegraphics[width=0.5\textwidth]{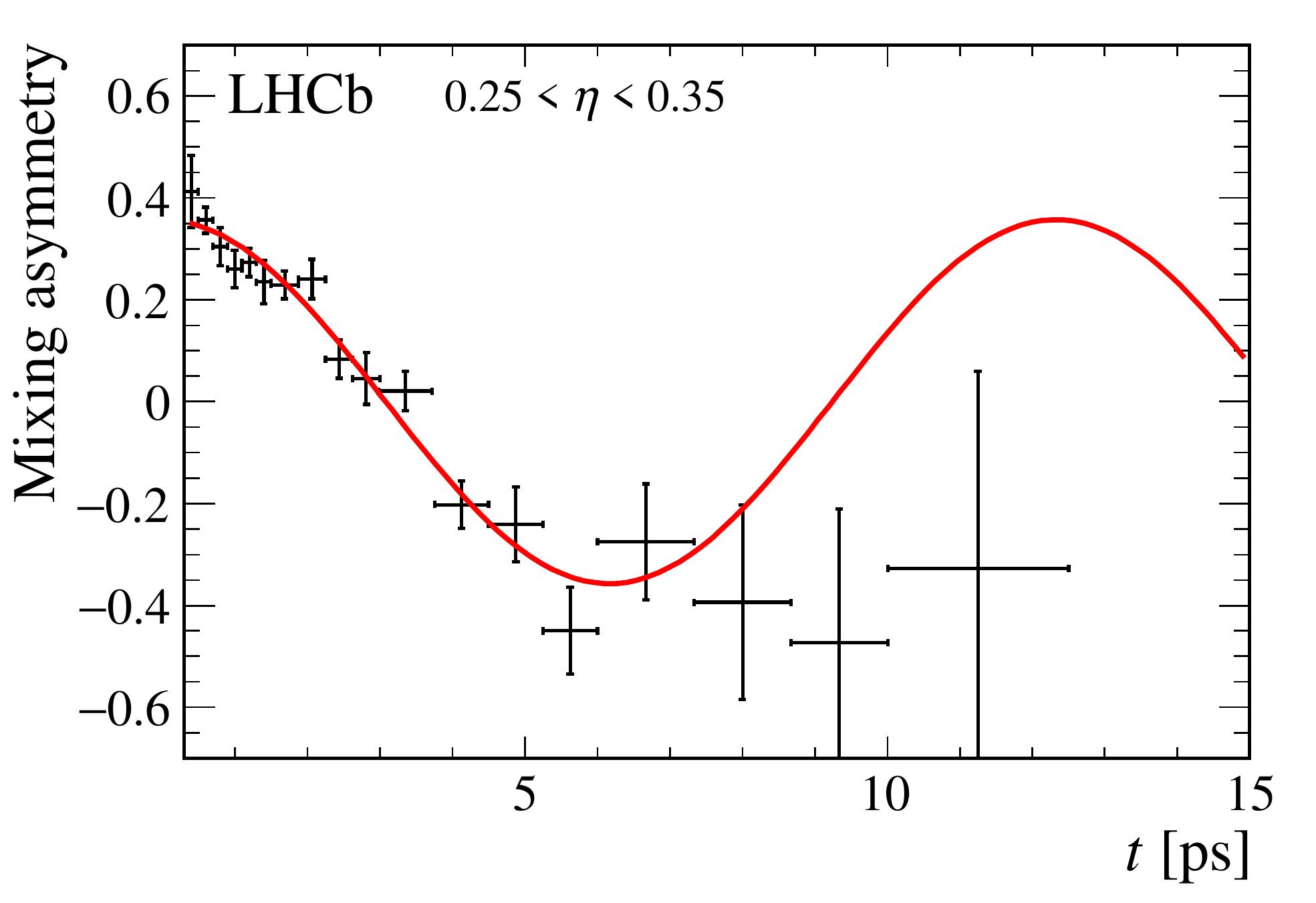}}
{\includegraphics[width=0.5\textwidth]{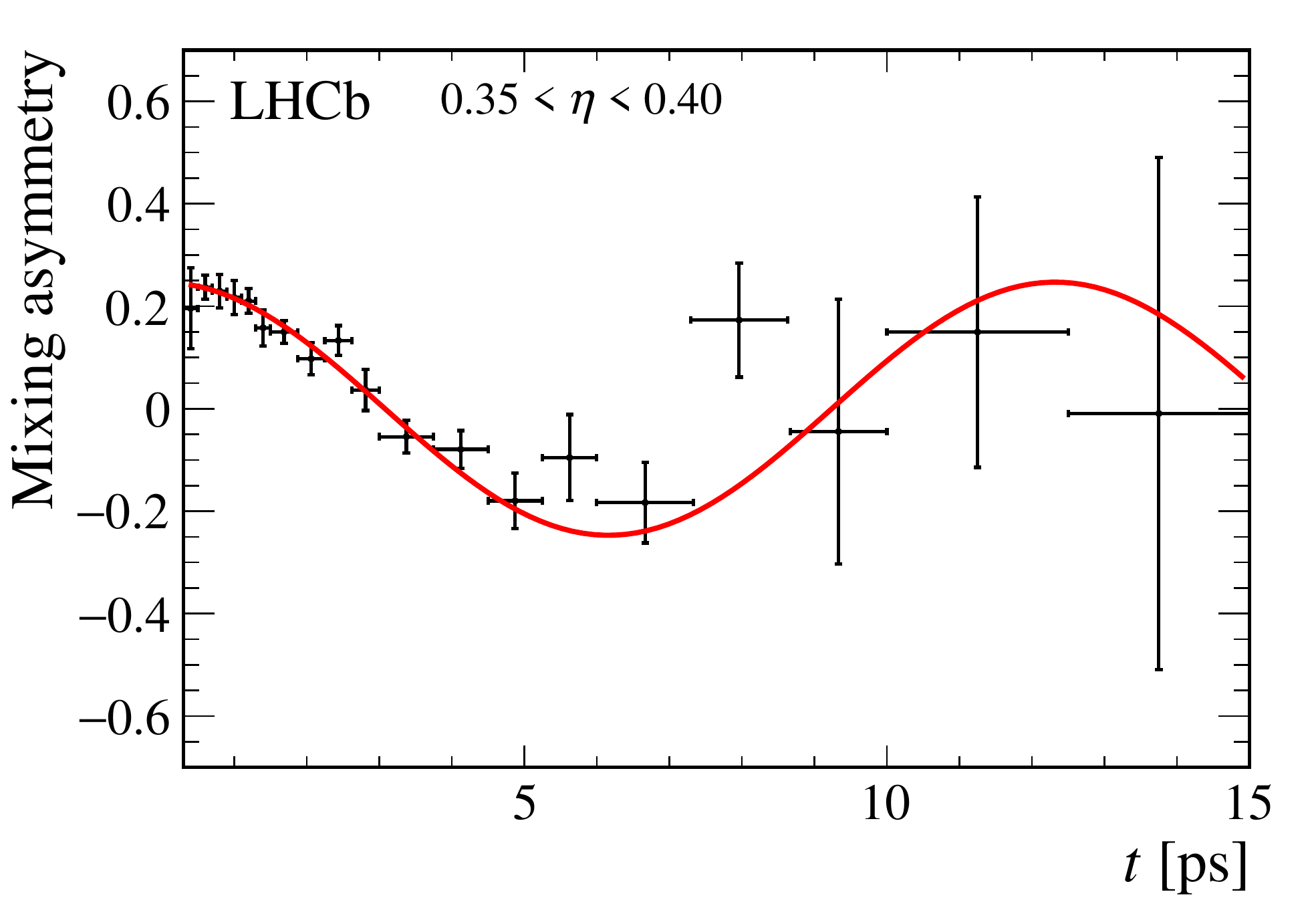}}
{\includegraphics[width=0.5\textwidth]{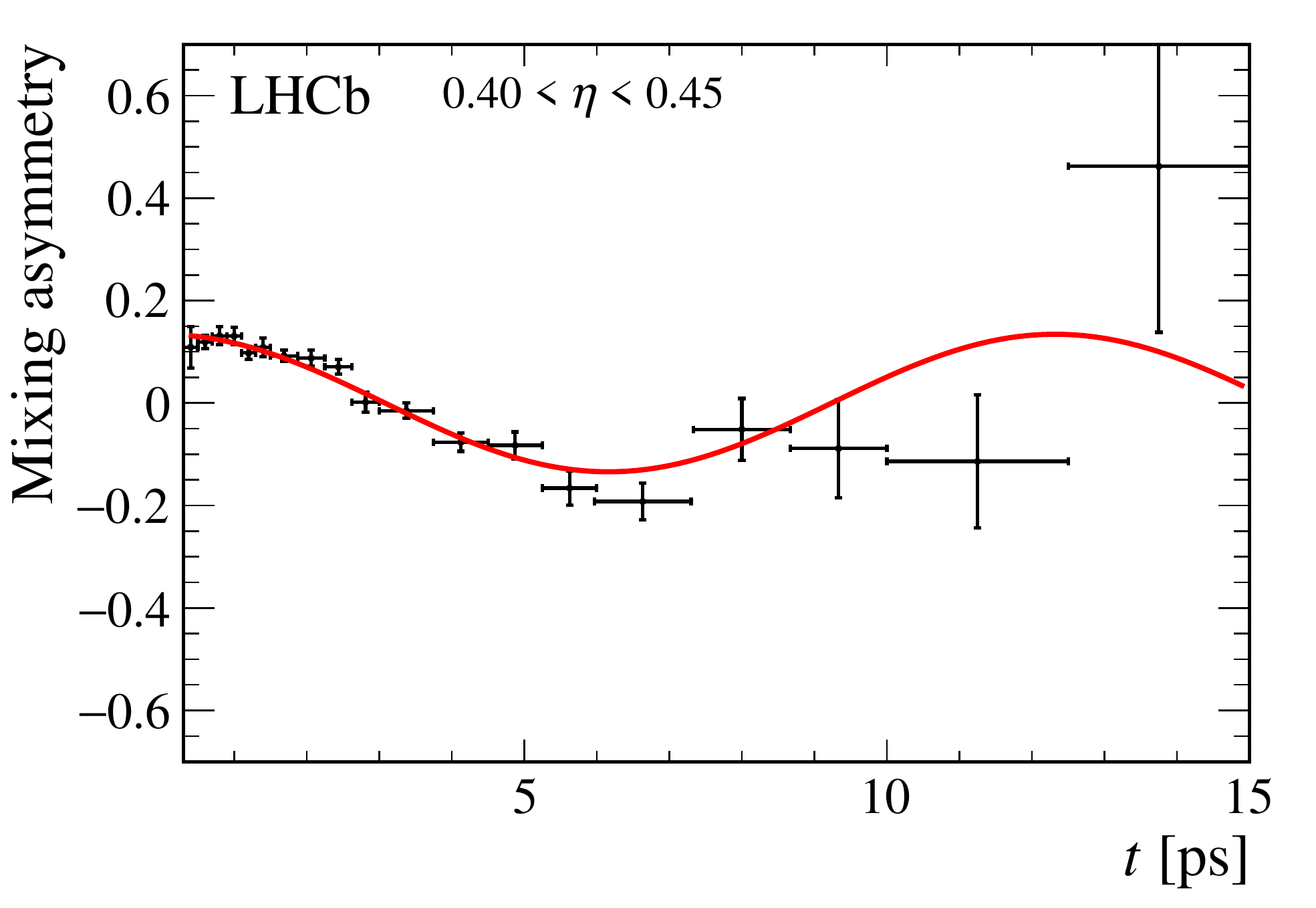}}
{\includegraphics[width=0.5\textwidth]{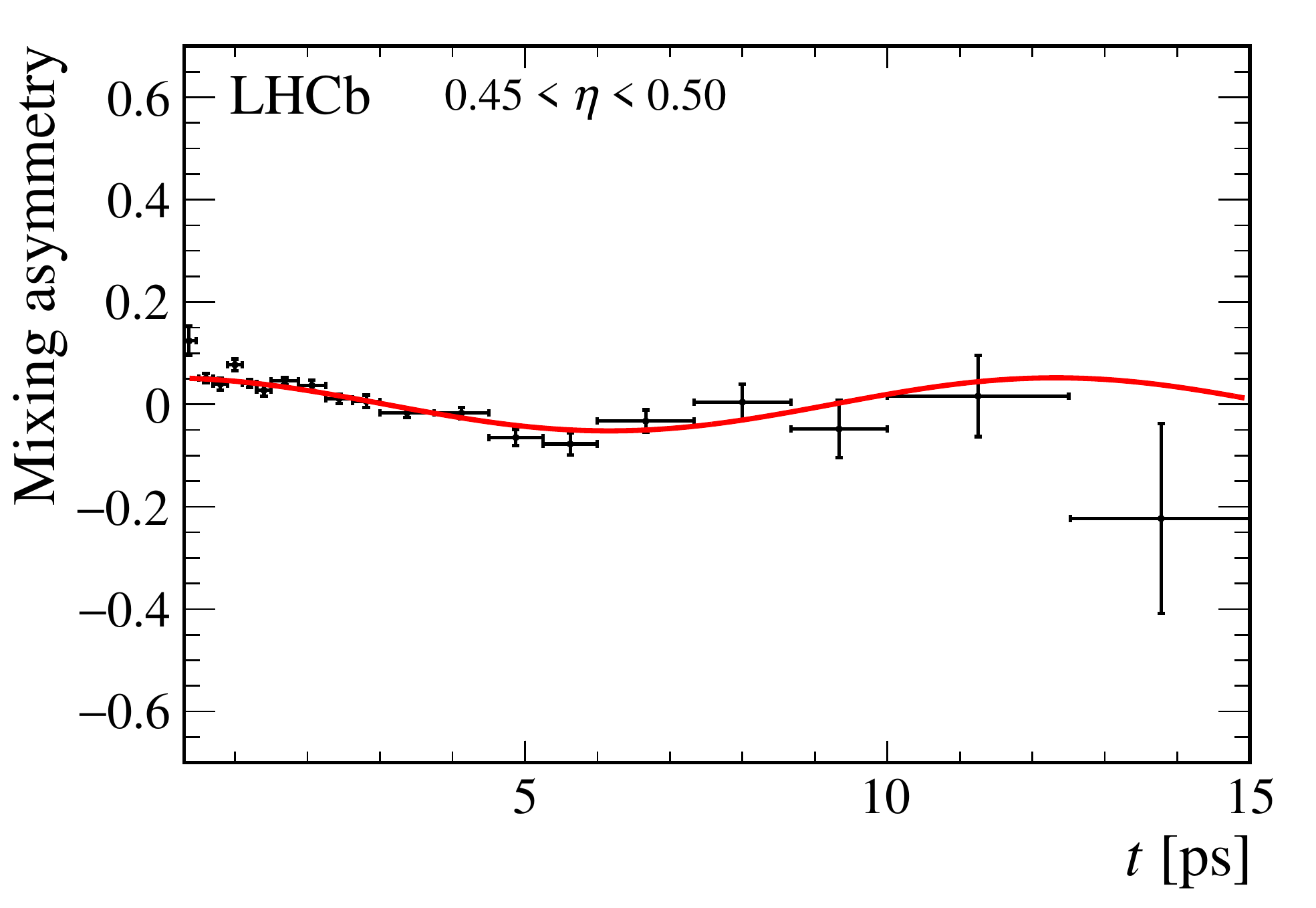}}
\caption{Mixing asymmetry in bins of mistag probability using the \SSpi tagger. 
}
\label{fig:AsymSSpi}
\end{figure}

\begin{figure}[tb]
{\includegraphics[width=0.5\textwidth]{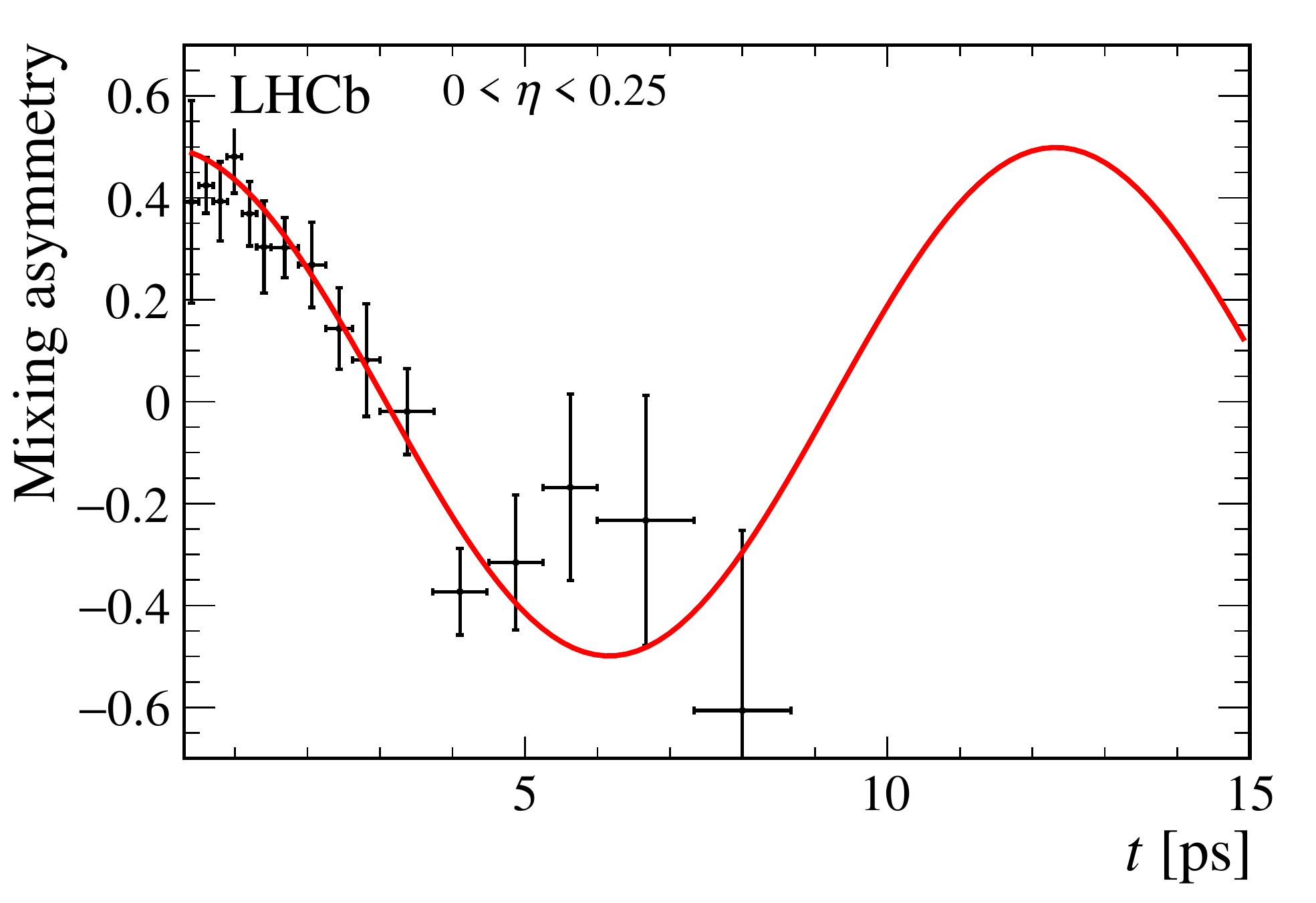}}
{\includegraphics[width=0.5\textwidth]{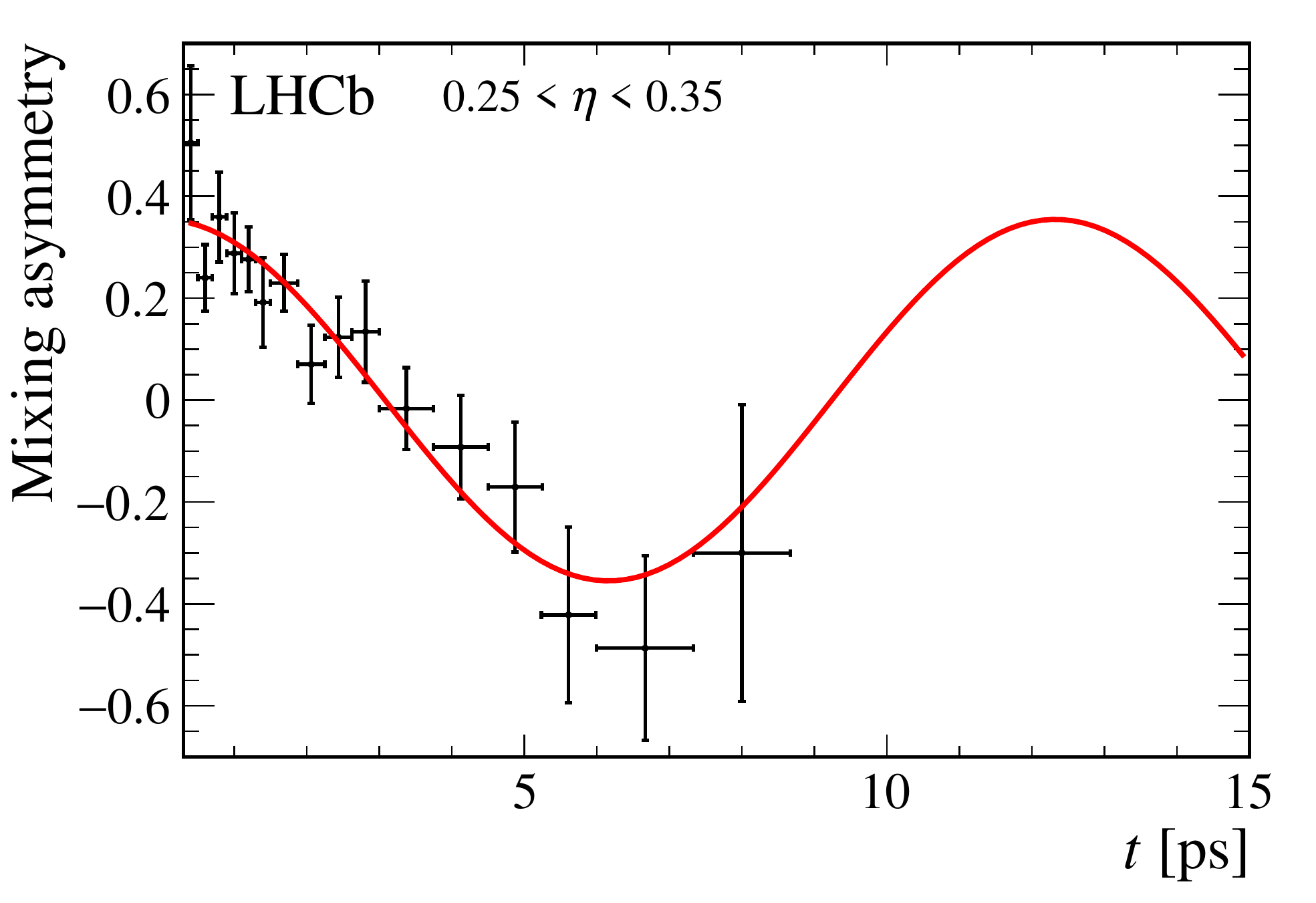}}
{\includegraphics[width=0.5\textwidth]{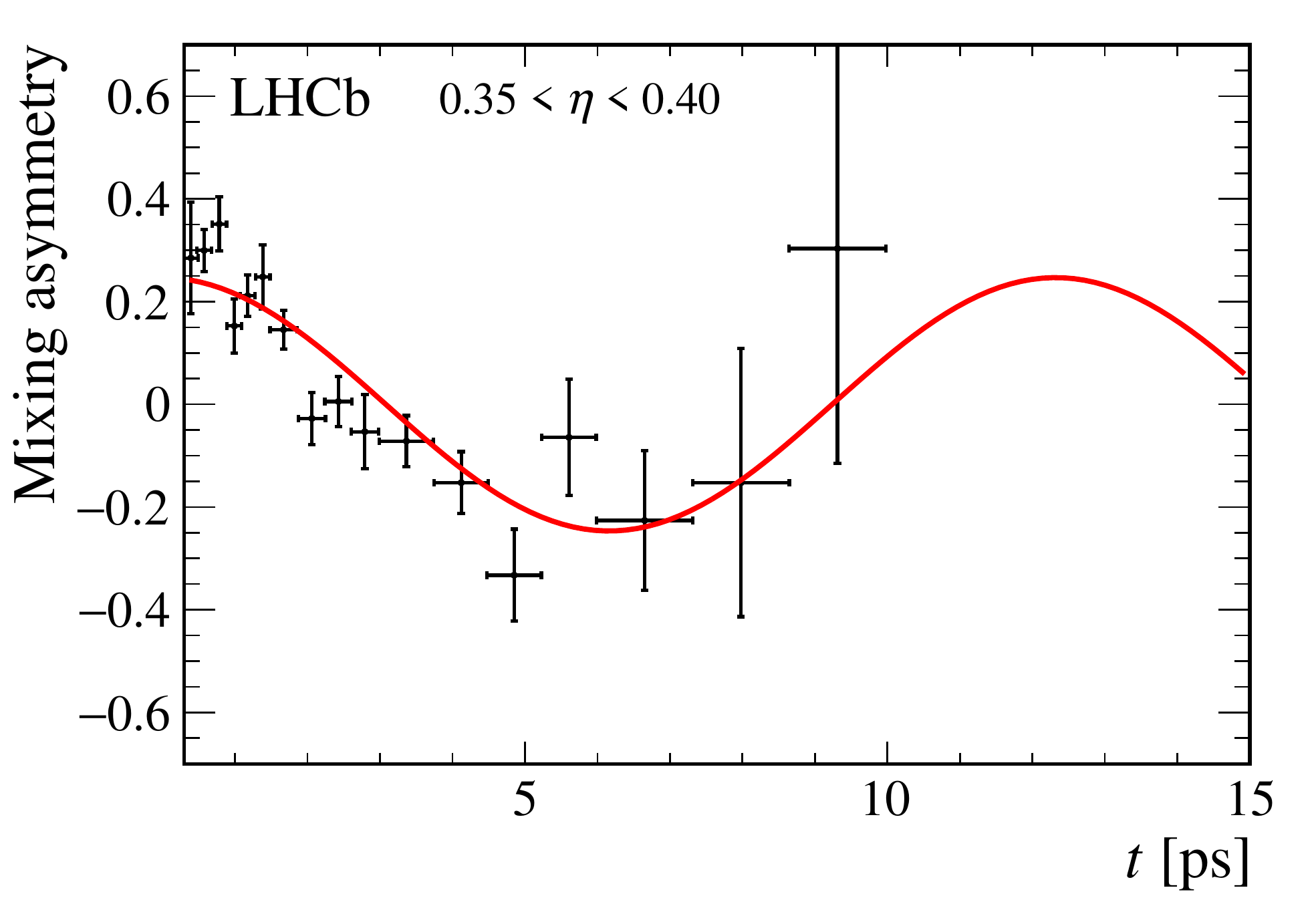}}
{\includegraphics[width=0.5\textwidth]{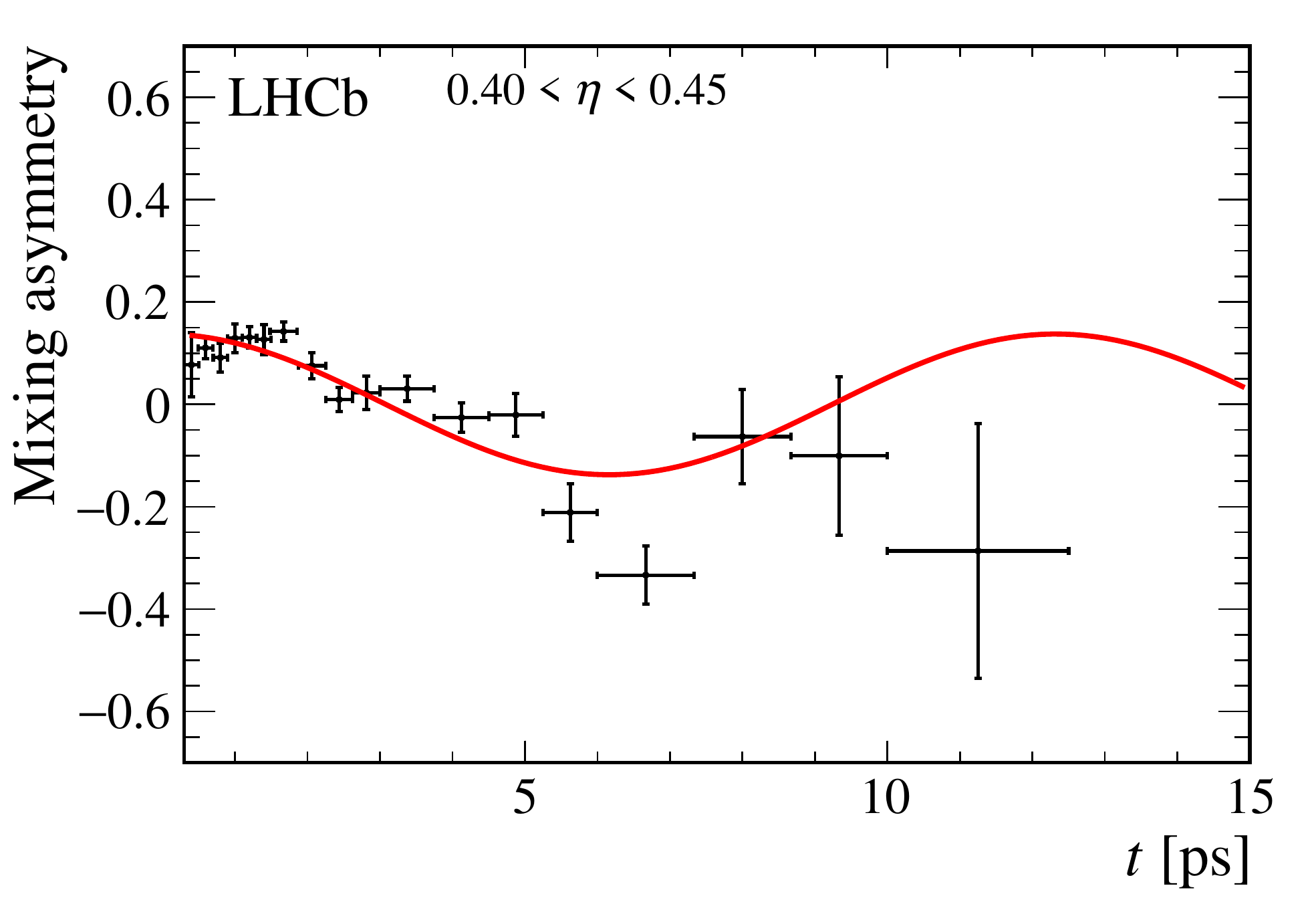}}
{\includegraphics[width=0.5\textwidth]{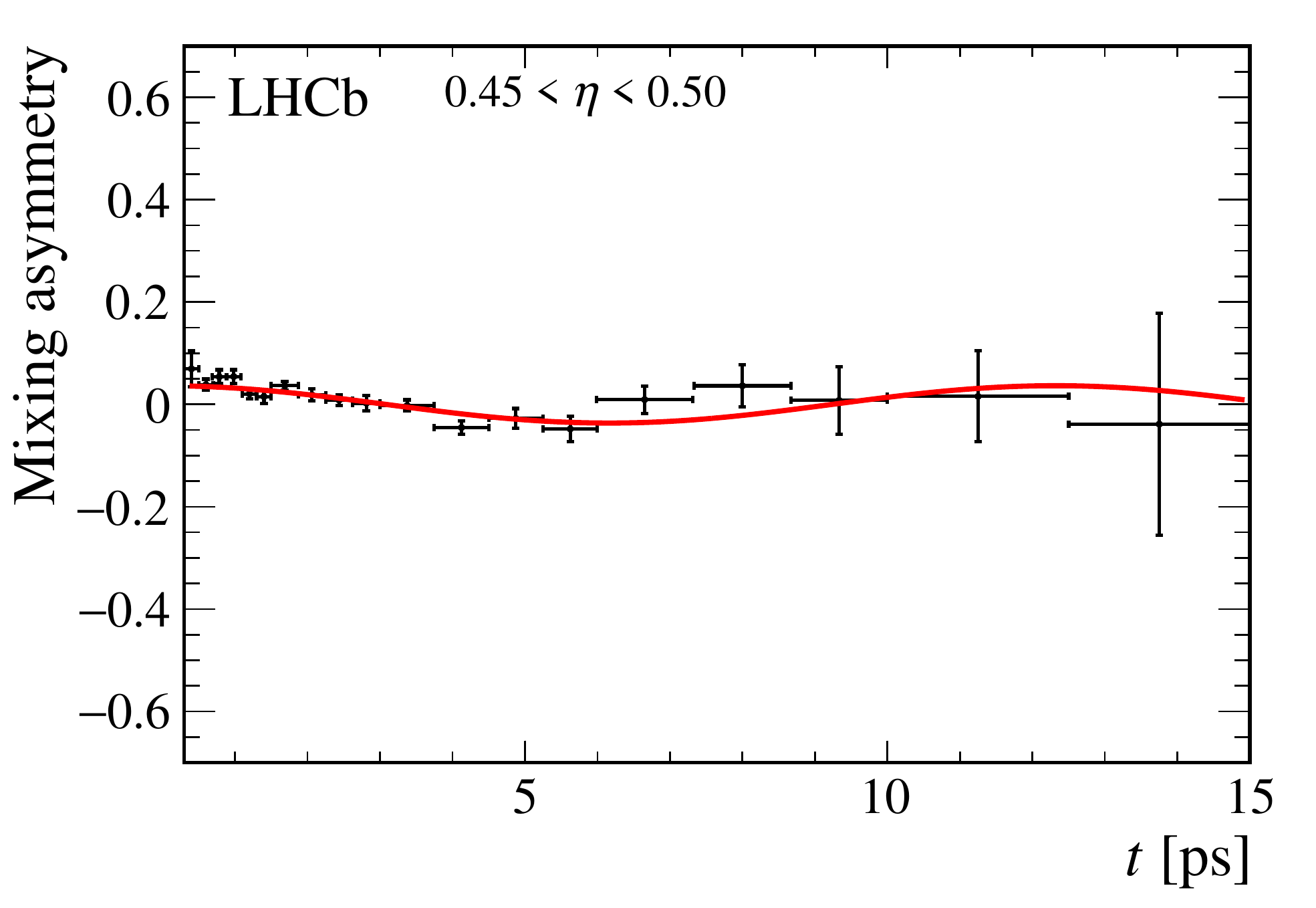}}
\caption{Mixing asymmetry in bins of mistag probability using the \SSp tagger.%
}
\label{fig:AsymSSp}
\end{figure}

The fit of the decay-time distribution is repeated after dividing events 
into bins of predicted mistag probability.
The distribution of $\eta$ and the dependence  
of the measured mistag fraction on $\eta$ are shown in Fig.~\ref{fig:etaSSpi} 
with the linear fits superimposed, demonstrating the expected linearity.
In Figs.~\ref{fig:AsymSSpi} and \ref{fig:AsymSSp} the time-dependent mixing asymmetries
\mbox{$A=\frac{N^{\mathrm{unmix}}-N^{\mathrm{mix}}}{N^{\mathrm{unmix}}+N^{\mathrm{mix}}}$}
are shown for each of the five bins.

\subsection{The SScomb tagger}
Even though a given tagging particle can be selected by only one of the \SSpi 
or the \SSp taggers, 
both taggers may find a candidate track in the same event. 
About 50\% of the candidates tagged by \SSpi are also tagged by \SSp, 
and 80\% of the candidates tagged by \SSp are also tagged by \SSpi. 
When both taggers provide a decision, they are combined into a single 
decision.
Since the correlation between the \SSpi and \SSp decisions, and between their 
mistag probabilities, is found to be small,
it is neglected when combining them using the following formulae 
\begin{equation}
 p(b) = \prod_i \bigg(\frac{1+d_i}{2} - d_i(1-\eta_i) \bigg), \qquad \qquad 
 p(\overline{b}) = \prod_i \bigg(\frac{1-d_i}{2}+d_i(1-\eta_i)\bigg),
\end{equation}
where p(\bquark) and p(\bquarkbar) are the probabilities that the signal \B meson contains a \bquark or a \bquarkbar  
quark respectively, and $d_i$ is the tagging decision of the tagger $i=$\SSpi, \SSp.
The normalized probabilities are
\begin{equation}
P(\overline{b}) = \frac{p(\overline{b})}{p(\overline{b}) + p(b)},  \qquad \qquad P(b) = 1 - P(\overline{b}).
\end{equation}
For $P(\overline{b}) > P(b)$ the combined tagging decision is $d=+1$ 
and the final mistag probability is $\eta = P(b)$.
Otherwise, the combined tagging decision 
and the mistag probability are $d=-1$ and $\eta=P(\overline{b})$.

The combination procedure, which assumes no correlation,
is validated by checking the combined mistag probability a posteriori.
Assuming a linear relation between the predicted mistag probability
and the true mistag fraction, 
the calibration parameters in the overlapping sample give 
$(\overline{p}_0 - \langle \eta \rangle)= 0.010 \pm 0.005$ 
and $(\overline{p}_1-1)= 0.01 \pm 0.08$.
The calibration is repeated on the sample of all \Bd candidates tagged 
by the \SScomb tagger, and 
the calibration parameters derived from the unbinned likelihood fit 
with the PDF of Eq.~\ref{eqn:Timefit}, reported in 
Table~\ref{tab:final_table}, demonstrate its validity.
The performance of \SScomb is reported in Table~\ref{tab:TagPowerSS}.
The total tagging power obtained by the combined algorithm is
$(2.11\pm 0.11)\%$, a relative increase of 25\% compared to that 
provided by the \SSpi tagger alone.

A higher tagging power can be obtained from the combination of the SScomb tagger 
with the OS tagger. The OS tagger is the combination of various 
OS tagging algorithms using electrons and muons from semileptonic decays of \bquark 
hadrons, kaons from $b \to c \to s$ decay chains and the inclusive 
reconstruction of a secondary vertex of the decay products of the 
opposite side \bquark hadron.
The SS and OS taggers are found to be uncorrelated, so their
combination follows the same procedure as the combination of \SSpi and \SSp into \SScomb.
The calibration of the combined mistag probability is verified a posteriori
with a fit of the decay-time distribution of the \Bd candidates.
For \BdDpi decays, the total tagging efficiency and the total tagging power
are $(84.48 \pm 0.26)\%$ and  $(5.14 \pm 0.15)\%$, respectively.
On the same sample, the use of the OS tagger only provides a tagging efficiency 
and a tagging power of $(37.95 \pm 0.15)\%$ and  $(3.52 \pm 0.17)\%$,
respectively.

\section{Validation and systematic uncertainties}
\label{sec:syst}
A possible dependence of the calibration parameters of the SS taggers on
properties of the event sample is checked by repeating the calibration 
after splitting the data 
according to the data-taking conditions (magnet polarity), global event 
properties (total number of reconstructed tracks, number of primary vertices) 
or according to the kinematic properties of the \Bd meson 
(transverse momentum, pseudorapidity and azimuthal angle).
The average mistag probability has a weak dependence on the number of tracks 
in the event. On the other hand, it decreases as a function of the 
transverse momentum
since the number of random tracks decreases at high $p_T^B$.
The tagging efficiency is nearly constant for pions, while the requirement
on proton identification reduces the number of proton candidates at high 
$p_T^B$.
A similar dependence is present versus the pseudorapidity of the \Bd meson. 
Since the average mistag fraction and the $p_0$ parameter 
decrease with increasing $p_{\rm T}^\Bd$, the calibration remains valid in 
all subsamples, with variations below two standard deviations.

The portability of the mistag calibration, from the training data sample to other data 
samples and other \Bd decay modes, is validated using an independent sample
of \BdDpi decays collected at \sqs = 7\tev (corresponding to an integrated luminosity of 1\invfb) and a sample of \BdToKpi decays
collected at \sqs = 8\tev (correspoding to an integrated luminosity of 2\invfb).
The same selection criteria and fitting procedure as described above are used for the 
\BdDpi validation sample at \sqs = 7\tev.
The calibration parameters for the \SSpi, \SSp, and \SScomb taggers
determined from an unbinned maximum likelihood fit to the decay-time distribution 
are compatible with those derived in the 8\tev sample. 
Consistent values of tagging power are found for all taggers. 

The selection criteria and the mass model for the \BdToKpi candidates are described in Ref.~\cite{LHCb-PAPER-2013-040}.
The decay-time acceptance is parametrized using cubic splines with six nodes,
whose positions are fixed and whose coefficients are free in the fit. 
The decay-time resolution is described by a Gaussian function with parameters 
determined from simulation.
The parameters shown in Table~\ref{tab:KpiCalib} demonstrate a good
portability of the mistag calibration, with
$\overline{p}_0 - \langle \eta \rangle \approx 0$ 
and $\overline{p}_1 - 1 \approx 0$ as expected.   
A lower tagging power is measured in this channel, giving
$(1.06 \pm 0.09 )$\%, $(0.42 \pm 0.06 )$\%, and 
$(1.37 \pm 0.13 )$\% for \SSpi, \SSp and \SScomb, respectively,
as expected from the lower average \pt of the selected \Bd candidates. 

\begin{table}[tb]
\centering
\caption{Calibration parameters for the \BdToKpi decay sample. 
Uncertainties are statistical only.}
 \resizebox{\textwidth}{!}{ 
 \begin{tabular}{ccccccc}
Tagger &$\langle \eta \rangle$& $\overline{p}_0$ & $\overline{p}_1$ & $\Delta p_0$ & $\Delta p_1$ & $A_{\mathrm{tag}}$\\ \hline
\SSpi  &$0.456$  & $0.452 \pm 0.003$ & $1.06 \pm 0.09$ & $\phantom{+}0.0053 \pm 0.0042$ & $0.047 \pm 0.115$ & $-0.009 \pm 0.008$ \\
\SSp   & $0.467$ & $0.459 \pm 0.004$ & $0.80 \pm 0.14$ & $-0.0138 \pm 0.0051$ & $0.025 \pm 0.141$ & $\phantom{+}0.008 \pm 0.009$ \\
\SScomb& $0.452$ & $0.457 \pm 0.003$ & $0.94 \pm 0.07$ & $-0.0034 \pm 0.0040$ & $0.079 \pm 0.086$ & $\phantom{+}0.007 \pm 0.007$ \\
 \end{tabular}
         }
\label{tab:KpiCalib}
\end{table}

Several sources of systematic uncertainties on the calibration parameters are studied
and the associated uncertainties are reported in Table~\ref{tab:systematics}. 
Uncertainties related to the mass model and background unfolding procedure are 
assessed by repeating the calibration replacing the sWeights derived in the fit 
to the mass distribution of all \Bd candidates by the sWeights derived after 
restricting the sample to tagged \Bd candidates.
In a second test, the signal mass model is replaced by a Hypatia 
function~\cite{Ipatia} convolved with a Gaussian function.
The sum in quadrature of the variations of the calibration parameters 
observed in 
the two tests is taken as uncertainty on the mass model.

Uncertainties related to the decay-time acceptance model are assessed by 
changing the number of nodes in the cubic splines from six to nine and 
are found to be  negligible.
A negligible uncertainty is associated to the decay-time resolution model.
The mistag model uncertainties are assessed by comparing the calibration 
parameters derived in the nominal fit and those derived in fits with the 
mistag probability binned in categories.
Five, seven and nine bins are tested and the largest observed variation of the 
parameters is taken as a systematic uncertainty.
Differences between the results of the two implementations of the 
time-dependent fit 
are due to the dependence of the mistag probability on the decay time.
Pseudoexperiments are generated where the mistag probability has 
the same dependence on time as in data and are fitted with the two approaches. 
The difference in parameters is similar to or smaller than that observed 
in data.

Uncertainties related to neglecting $\Delta \Gamma_d$ and possible 
\CP violation in the \BdDpi decays in the decay-time fit,
are studied by performing pseudoexperiments in which changes associated 
with the parameter under study are incorporated in the generation and neglected 
in the subsequent fit. 
Terms proportional to the relevant \CP parameters are added to the PDF 
in Eq.~\ref{eqn:Timefit} and the values of the parameters are taken 
from Ref.~\cite{HFAG}.
The associated systematic uncertainties are taken to be the changes in the 
calibration parameters with respect to perfect calibration
($\overline{p}_0=\langle \eta \rangle$, $\overline{p}_1$=1), used in the generation.
Uncertainties related to the variation of $A_{\mathrm{prod}}$ and $a_{\mathrm{sl}}^d$,
which are fixed in the decay-time fit, are evaluated with pseudoexperiments
where the parameters are varied within their uncertainties.
The uncertainties are determined in the \SSpi configuration 
and attributed to both taggers. A breakdown of the systematic uncertainties 
related to the decay-time model is shown in Table~\ref{tab:sysmodel}.

 \begin{table}[tb]                                                          
 \centering    
 \caption[]{Systematic uncertainties on the calibration parameters of \SSpi, \SSp and \SScomb taggers. 
   The total systematic uncertainty is the squared sum of all contributions. A dash indicates a value
   negligible with respect to the quoted precision.}
\begin{tabular}{ccccccc}                                                                          
Tagger &Source & $\sigma (\overline{p}_0)$ & $\sigma (\overline{p}_1)$ & $\sigma (\Delta p_0)$ & $\sigma (\Delta p_1)$ &  $\sigma (A_{\mathrm{tag}})$ \\ \hline

\multirow{4}{*}{\SSpi} & mass model &  -- & -- & -- & $0.001$ & -- \\  
&mistag model& $0.001$ & $0.01$ & $0.0002$ & $0.007$ & -- \\
&decay model & 0.001 & 0.01 & 0.0016 & 0.012 & 0.007 \\ 
\cline{2-7}
&total & $0.001$ & $0.01$ & $0.0016$ & $0.014$ & $0.007$\\ 
\hline
\multirow{4}{*}{\SSp} & mass model &  -- & -- & $0.0002$ & $0.004$ & --  \\
&mistag model & 0.001 & 0.02 & -- & 0.014 & 0.001 \\  
&decay model & 0.001 & 0.01 & 0.0016 & 0.012 & 0.007 \\ 
\cline{2-7}
&total & $0.001$ & $0.02$ & $0.0016$ & $0.019$ & $0.007$ \\     
\hline
\multirow{4}{*}{\SScomb} & mass model & -- & -- & $0.0008$ & $0.005$ & --  \\  
&mistag model & 0.002 & 0.02 & 0.0004 & 0.010 & 0.001 \\  
&decay model & 0.001 & 0.01 & 0.0016 & 0.012 & 0.007 \\ 
\cline{2-7}
&total & $0.002$ & $0.02$ & $0.0018$ & $0.017$ & $0.007$ \\  
\end{tabular}    
\label{tab:systematics}  
\end{table}                       

\begin{table}[tb]
  \centering
  \caption{Systematic uncertainties related to the decay-time model. 
A dash indicates a value negligible with respect to the quoted precision.}
  \begin{tabular}{cccccc}
 Source & $\sigma(\overline{p}_0)$ & $\sigma (\overline{p}_1)$ & $\sigma(\Delta p_0)$ & $\sigma (\Delta p_1)$ & $\sigma(A_{\mathrm{tag}})$  \\

    \hline
    $\Delta \Gamma$ & 0.00013 & -- & -- & -- & 0.001 \\
    $A_{\mathrm{prod}}$ & 0.00002 & -- & -- & -- & 0.007 \\
    $a_{\mathrm{sl}}^{\mathrm{d}}$ & -- & -- & -- & -- & -- \\
    \CP violation & 0.00124 & 0.01 & 0.0016 & 0.012 & 0.002 \\
    \hline
    total & 0.001 & 0.01 & 0.0016 & 0.012 & 0.007 \\
  \end{tabular}
  \label{tab:sysmodel}
\end{table}

\section{Conclusion}
Two new same-side algorithms are developed to determine the production 
flavour of \Bd mesons using pions and protons from the hadronization 
process.
This is the first time that protons are used to identify the flavour of
a \Bd meson.
The algorithms are optimized and calibrated on data using
\BdDpi decays.
The calibration parameters of the taggers are reported in Table~\ref{tab:final_table}.
The efficiency and mistag probability of the taggers depend on the kinematic 
properties of the \Bd decay mode under study.
Estimated mistag probabilities match the true mistag fraction 
throughout the phase space. 
The new \SSpi tagger provides a tagging power that is greater by 60\% 
relative to the previous algorithm using pions, employed 
in Ref.~\cite{LHCb-PAPER-2015-004}.
Adding the combination of the two new algorithms to the existing OS taggers
provides a relative increase of the total tagging power of about 40\%.

\section*{Acknowledgements}
\noindent We express our gratitude to our colleagues in the CERN
accelerator departments for the excellent performance of the LHC. We
thank the technical and administrative staff at the LHCb
institutes. We acknowledge support from CERN and from the national
agencies: CAPES, CNPq, FAPERJ and FINEP (Brazil); NSFC (China);
CNRS/IN2P3 (France); BMBF, DFG and MPG (Germany); INFN (Italy); 
FOM and NWO (The Netherlands); MNiSW and NCN (Poland); MEN/IFA (Romania); 
MinES and FASO (Russia); MinECo (Spain); SNSF and SER (Switzerland); 
NASU (Ukraine); STFC (United Kingdom); NSF (USA).
We acknowledge the computing resources that are provided by CERN, IN2P3 (France), KIT and DESY (Germany), INFN (Italy), SURF (The Netherlands), PIC (Spain), GridPP (United Kingdom), RRCKI and Yandex LLC (Russia), CSCS (Switzerland), IFIN-HH (Romania), CBPF (Brazil), PL-GRID (Poland) and OSC (USA). We are indebted to the communities behind the multiple open 
source software packages on which we depend.
Individual groups or members have received support from AvH Foundation (Germany),
EPLANET, Marie Sk\l{}odowska-Curie Actions and ERC (European Union), 
Conseil G\'{e}n\'{e}ral de Haute-Savoie, Labex ENIGMASS and OCEVU, 
R\'{e}gion Auvergne (France), RFBR and Yandex LLC (Russia), GVA, XuntaGal and GENCAT (Spain), Herchel Smith Fund, The Royal Society, Royal Commission for the Exhibition of 1851 and the Leverhulme Trust (United Kingdom).

\addcontentsline{toc}{section}{References}
\setboolean{inbibliography}{true}

\bibliographystyle{LHCb}
\bibliography{main,LHCb-PAPER,LHCb-DP}

\newpage
  
\centerline{\large\bf LHCb collaboration}
\begin{flushleft}
\small
R.~Aaij$^{40}$,
B.~Adeva$^{39}$,
M.~Adinolfi$^{48}$,
Z.~Ajaltouni$^{5}$,
S.~Akar$^{6}$,
J.~Albrecht$^{10}$,
F.~Alessio$^{40}$,
M.~Alexander$^{53}$,
S.~Ali$^{43}$,
G.~Alkhazov$^{31}$
% ...

{\footnotesize \it
% ...
$ ^{5}$Clermont Universit{\'e}, Universit{\'e} Blaise Pascal, CNRS/IN2P3, LPC, Clermont-Ferrand, France\\
$ ^{6}$CPPM, Aix-Marseille Universit{\'e}, CNRS/IN2P3, Marseille, France\\
% ...
$ ^{10}$Fakult{\"a}t Physik, Technische Universit{\"a}t Dortmund, Dortmund, Germany\\
% ...
$ ^{31}$Petersburg Nuclear Physics Institute (PNPI), Gatchina, Russia\\
% ...
$ ^{39}$Universidad de Santiago de Compostela, Santiago de Compostela, Spain\\
$ ^{40}$European Organization for Nuclear Research (CERN), Geneva, Switzerland\\
% ...
$ ^{43}$Nikhef National Institute for Subatomic Physics, Amsterdam, The Netherlands\\
% ...
$ ^{48}$H.H. Wills Physics Laboratory, University of Bristol, Bristol, United Kingdom\\
% ...
$ ^{53}$School of Physics and Astronomy, University of Glasgow, Glasgow, United Kingdom\\
% ...
}
\end{flushleft}                                                     
\end{document}